%% file: ms.tex
\documentclass[apj]{emulateapj}
\input{Definitions.tex}
\newcommand{\taues}{\tau_{\text{esc}}}
\newcommand{\tauesc}{\taues}
\newcommand{\esc}{\text{esc}}

\newcommand{\taucool}{\tau_{c}}
\newcommand{\GV}{\text{GV}}

\newcommand{\Xesc}{X_{\text{esc}}}
\newcommand{\Xdisc}{X_{\text{disc}}}
\usepackage{amsmath}

\begin{document}

\title{What can we really learn from positron flux 'anomalies'?}

\author{Boaz Katz\altaffilmark{1}, Kfir Blum\altaffilmark{1}, Jonathan Morag\altaffilmark{1} and
Eli Waxman\altaffilmark{1}} \altaffiltext{1}{Physics Faculty, Weizmann Institute of Science, Rehovot, 76100 Israel}

\begin{abstract}
We present a critical analysis of the observational constraints on, and of the theoretical modeling of, aspects of cosmic
ray (CR) generation and propagation in the Galaxy, which are relevant for the interpretation of recent positron and
anti-proton measurements. We give simple, analytic, model independent expressions for the secondary $\bar{p}$ flux, and
an upper limit for the secondary $e^+$ flux, obtained by neglecting $e^+$ radiative losses, $e^+/(e^+ + e^-)<0.2\pm0.1$
up to $\sim 300$ GeV. These expressions are completely determined by the rigidity dependent grammage, which is measured
from stable CR secondaries up to $\sim150$ GeV/nuc, and by nuclear cross sections measured in the laboratory. $\bar{p}$
and $e^+$ measurements, available up to $\sim100$ GeV, are consistent with these estimates, implying that there is no
need for new, non-secondary, $\bar{p}$ or $e^+$ sources. The radiative loss suppression factor $f_{s,e^+}$ of the $e^+$
flux depends on the $e^+$ propagation in the Galaxy, which is not understood theoretically. A rough, model independent
estimate of $f_{s,e^+}\sim 1/3$ can be obtained at a single energy, $\vep\sim20$ GeV, from unstable secondary decay and
is found to be consistent with $e^+$ measurements, including the positron fraction measured by PAMELA. We show that
specific detailed models, that agree with compositional CR data, agree with our simple expressions for the $e^+$ and
$\bar{p}$ flux, and that the claims that the positron fraction measured by PAMELA requires new primary $e^+$ sources are
based on assumptions, that are not supported by observations. If PAMELA results are correct, they suggest that
$f_{s,e^+}(\vep)$ is slightly increasing with energy, which provides an interesting constraint on CR propagation models.
We argue that measurements of the $e^+$ to $\bar{p}$ ratio are more useful for challenging secondary production models
than the $e^+/(e^+ + e^-)$ fraction.
\end{abstract}

% -------------------------- End of abstract -----------------------


% -----------------------------------------------------------------------
% --------------------------  Sec 1: INTRODUCTION -----------------------
% -----------------------------------------------------------------------

\section{Introduction}
\label{sec:Introduction}

Recently, the positron \citep{Adriani09a} and anti-proton \citep{Adriani09b} Cosmic Ray (CR) fluxes were measured to high
energies, $\vep\sim 100\GeV$, with unprecedented accuracy by the PAMELA satellite-borne experiment. An anomalous positron
overabundance, compared to the expected abundance of secondary positrons produced by interactions of CRs and ambient
(inter-stellar) nuclei, was reported to exist in the energy range $1.5\GeV<\vep <100\GeV$ based on the reported ratio of
the $e^+$ flux to the sum of the $e^-$ and $e^+$ fluxes \citep{Adriani09a}. This reported overabundance was widely
claimed to necessarily imply the existence of primary $e^+$ sources \citep[e.g.][]{Adriani09a,Morselli08}. Both
astrophysical objects \citep[see e.g.][and references within]{Profumo08} and annihilation of dark matter particles in our
Galaxy \citep[see e.g.][and references within]{Meade09} were suggested as possible sources.

Assuming that the PAMELA measurements of the positron fraction are correct \citep[see e.g.][for cautionary
notes]{Schubnell09}, the robustness of the claim of the necessity of a primary $e^+$ source depends on the robustness of
the theoretical predictions for the secondary $e^+$ flux, and on the reliability of the estimate of the $e^+ + e^-$ flux,
theoretical or observational. As for the $e^+ + e^-$ flux, recent precise measurements are available [\citep{Abdo09,
Chang08} and also \citep{Aharonian08, Aharonian09}], especially in the high energy range $\vep\gtrsim 30\GeV$. As for the
theoretical predictions of the $e^+$ flux, these were obtained in the majority of recent publications by using detailed,
often numerical calculations of specific models, that are based on differing assumptions regarding CR generation and
propagation \citep[see e.g.][and references within]{Morselli08,Delahaye08}. The most commonly used, unestablished
assumption is that the CR propagation is described on large scales by an isotropic diffusion equation. Moreover, in most
cases the diffusion coefficient is assumed to be homogenous or to exhibit a simple distribution with energy independent
boundary conditions. In these models, the known relevant propagation properties are mixed with unestablished assumptions
and it is difficult to separate between robust and model dependent results.

In fact, a rather accurate estimate of the flux of any secondary CR resulting from CR-Inter Stellar Medium (ISM)
interactions, can be made without a detailed understanding of the CR propagation \citep{Gaisser92}. The spatial and
temporal dependence of the source functions for all such secondaries are the same, as they are proportional to the
product of the CR intensity and the ISM density (assuming that the primary CRs have a uniform composition). The ratio of
the local fluxes of two secondary particles should be equal to the measured ratio of their local production rates. Thus,
a measurement of one secondary specie (e.g. Boron), allows a prediction of all other secondaries, given that the
interaction cross sections are known. The main challenge in the analysis comes from the fact that CRs suffer significant
losses, which are different for different species, during their propagation. Nuclei suffer spallation and decay, while
positrons suffer energy losses due to Inverse Compton and Synchrotron emission. Anti-protons suffer some losses due to
annihilation and scattering.

Guided by these considerations, we present in this paper a critical analysis of the observational constraints on, and the
theoretical modeling of, aspects of CR generation and propagation in the Galaxy, which are relevant for the
interpretation of recent $e^+$ and $\bar p$ measurements. The plan of the paper, along with some of the principle points,
is as follows.

In \sref{sec:Grammage} we review the main observationally established properties of CR propagation. We focus on the
analysis of stable CR nuclei measurements and show that they can be accurately modeled using transparent physical
arguments. In particular, the losses due to spallation are addressed. The CR grammage, extracted from the stable CR
nuclei measurements, allows us to write down in \sref{sec:simple} analytic, model independent constraints for the
expected secondary $\bar p$ and $e^+$ fluxes, including an accurate estimate of the $\bar p$ flux and an upper limit for
the $e^+$ flux, obtained by neglecting the $e^+$ radiative losses. Anti-proton and $e^+$ measurements available up to
$\vep\sim 100$ GeV, including the PAMELA positron fraction measurement, are shown to be consistent with these estimates.

The $e^+$ flux measurement can be used to infer the suppression factor $f_{s,e^+}(\vep)$  due to radiative energy losses,
defined by the relation
\begin{equation}\label{eq:fDef}
n_{e^+}(\vep)=f_{s,e^+}(\vep)n_{e^+,\text{no losses}}(\vep),
\end{equation}
where $n_{e^+,\text{no losses}}$ is the denisty of positrons that would have been observed if positrons did not suffer
energy losses. Similarly to secondary nuclei and anti-protons, $n_{e^+,\text{no losses}}$ can be calculated directly
using the measured CR grammage. We show that the existing $e^-$ and $e^+$ data imply that $f_{s,e^+}(\vep)\sim
1/\text{few}$ in the PAMELA energy range, and is indicated to be slightly growing with energy.

In \sref{sec:losses} we discuss the measurements that allow an indirect estimate of $f_{s,e^+}(\vep)$, namely via
comparison to the suppression of the abundance of radioactive elements. We show that observations allow a rough, model
independent, estimate of $f_{s,e^+}\sim1/3$ at a single energy $\vep\sim20\GeV$. This estimate is shown to be consistent
with $e^+$ measurements, implying that (i) positrons are almost certainly entirely secondary, and (ii) $e^+$ measurements
at high energies $\vep\gtrsim 20$ GeV provide novel information on the temporal behavior of CR propagation that is not
currently accessible by other types of experiments.

Any propagation model that reproduces the CR compositional measurements discussed above will be subject to the
constraints we provided for the positron and anti-proton fluxes. In particular it should agree with our estimates for the
secondary anti-proton flux, should agree with the estimate for the secondary positron flux at $\vep\sim20\GeV$ and should
give positron fluxes lower than the upper limit we derived, at positron energies $\vep\gtrsim 20\GeV$.  In section
\sref{sec:OtherModels} we illustrate this point by briefly discussing specific, commonly used models for CR propagation,
focusing on the Leaky Box and disc+halo diffusion models. Details are given in appendix \sref{sec:models}. The equations
for stable nuclei and anti-protons in the Leaky Box model and thin-disc+halo diffusion models, are shown to be equivalent
to our equations, as long as the grammage is set to be equal to the measured grammage. We reproduce the results of
detailed calculations of anti-protons and positrons using simple analytical approximations for the grammage and energy
loss suppression factor. We note that similar simple analytic expressions for the expected fluxes of secondary particles
in these specific models can be found in the literature \citep[e.g.][]{Ginzburg76,Berezinskii90,Longair92,Ptuskin09}. We
highlight the main, frequently unestablished, assumptions leading to the claimed conclusion that secondary models fail to
reproduce the $e^+$ measurements.

Previous claims, that a primary $e^+$ source is necessary to explain the positron fraction observations, are discussed in
\S~\ref{sec:claim_prim}. We show that these claims are based on assumptions regarding the generation spectrum of primary
electrons and regarding the propagation of electrons and positrons, which are not supported by observations.

In section \sref{sec:SecModels} we discuss specific models, in which the positrons are of secondary origin, that were
suggested to explain recent $e^+$ abundance measurements \citep{Blasi09,Shaviv09,Cowisk09a}. Although these models do not
adopt the assumptions discussed in \S~\ref{sec:claim_prim}, which lead to a conflict with the PAMELA results, they do
adopt other assumptions, which lead to conflicts with constraints derived from other CR measurements. We emphasize,
though, that there is a variety of models that can be constructed, in which both $e^+$ and anti-protons are of secondary
origin, and which are consistent with the constraints arising from CR observations. The main point that should be
addressed in such models is the energy dependence of the $e^+$ suppression factor $f_{s,e^+}$ at energies exceeding
$\vep\gtrsim 20\GeV$, which is indicated by the PAMELA measurements to be slightly rising with energy.

Our results are discussed and summarized in \sref{sec:Discussion}.

Throughout the paper, we limit the discussion to relativistic $\vep>10$ GeV/nuc energies, since at sub GeV energies the theory of CRs becomes more complicated and uncertain. This is due to the presence of various effects, including
ionization losses, energy dependent spallation cross sections and charge dependent solar modulation etc. Some of the
observed effects are not well understood (e.g. solar modulation) and possible additional effects may exist (e.g.
reacceleration).

\section{CR Grammage}\label{sec:Grammage}
A complete coverage of the theoretical and observational efforts to understand CR origin and propagation is beyond the
scope of this paper. In this section and in section \sref{sec:losses} we emphasize what we believe are the main,
observationally confirmed properties of CRs that allow for predictions of secondary $e^+$ and $\bar p$ fluxes.

 Although the acceleration and propagation of
CRs in our Galaxy has been studied for many decades [for reviews see
\citet{Ginzburg69,Blandford87,Berezinskii90,Longair92} and more recently \citet{Strong07}], based on results of numerous
experiments, to date still only little is known. Current knowledge is essentially restricted to the following features:
\begin{enumerate}
\item A qualitative global picture. CRs are generated in the Galaxy and are confined by the non trivial magnetic field
of the Galaxy to long, rigidity dependent time scales ($\sim 10^7 \yr$ at $\vep\sim1\GeV$). The CR halo is probably
considerably thicker than the few-hundred-pc-thick gaseous disc. Indirect evidence (notably synchrotron emission from
nearby galaxies and gamma-rays from our Galaxy) suggest that the CR distribution is, to an order of magnitude, homogenous
throughout the halo.

\item
Some quantitative features of CR propagation, established by CR compositional measurements. The most important
quantitative feature relevant to the calculation of secondary particles is the rigidity dependent average amount of
column density (\textit{grammage}) traversed by CRs. Information on CR grammage is based on measurements of the relative
abundance of spallation generated secondary particles such as Li,Be and B or Sc,Ti and V up to energies $\vep\lesssim
150$ GeV/nuc. Limited constraints on the propagation time scales exist due to the observed suppression in the abundance
of unstable radioactive nuclei, especially $^{10}$Be at low energies, $\vep\sim 100$ MeV/nuc.
\end{enumerate}

As CRs traverse the interstellar medium they suffer spallation. This leads to a few features that are
observed in the local CR nuclei spectrum and composition:
\begin{itemize}
\item An over abundance with respect
to the solar composition of chemical elements that are produced in the spallation interactions. For some elements like Li,Be,B and Sc,V,Ti, this over abundance is by a few orders of
magnitude, and thus these are most probably purely secondary particles.
\item A suppression of the flux of particles that were lost due to spallation.  This effect grows with the total
cross section for spallation, which in turn grows with the particles mass and reaches a factor of a few for Iron.
\end{itemize}

The ``optical depth" for spallation of a particle is proportional to the amount of grammage surpassed by the particle,
irrespective of the time it took the particle to accumulate this grammage or to the densities through which it traversed.

In this section, we discuss the measured amount of grammage deduced from the compositional analysis of stable nuclei.
This allows us to derive in \sref{sec:simple} model independent expressions for the secondary $\bar p$ flux, and an upper
limit for the $e^+$ flux, obtained by neglecting $e^+$ radiative losses. For clarity we begin in \sref{sec:Simplest} by
considering secondary particles that do not suffer losses. The definition of the CR grammage and the use of secondary
particles to deduce it's value are discussed. Spallation losses are addressed in \sref{sec:SimpleSpal}, where we discuss
the observations and the measured grammage values. The analysis of measurements of radioactive unstable nuclei are
discussed in \sref{sec:losses}, where they are used to derive constraints on the $e^+$ radiative losses suppression
factor.

\subsection{Stable secondaries without spallation losses}\label{sec:Simplest}

Consider hypothetical CR secondaries that suffer no losses at all. The local density of such CRs at a given rigidity
will be proportional to their local generation rate under the following assumptions:

\begin{equation}\label{eq:Conditions}\end{equation}
\begin{enumerate}
\item secondary particles with the same rigidity propagate through the ISM in the same way (diffusively or otherwise),
\item the rigidity of the products equals the rigidity of the primary,
\item the energy of secondary particles does not change during propagation,
\item the composition (but not necessarily flux, spectrum or target density) of CRs is uniform throughout the region in which most of the secondaries observed here are produced and during the time they were produced.
\end{enumerate}
Under the above conditions, for any two stable CRs A and B at a given rigidity, the following equation holds:
\begin{equation}\label{eq:SecondaryRatioEquationNS}
\frac{n_A}{n_B}=\frac{Q_{A}}{Q_{B}},
\end{equation}
where $Q_{i}$ is the local production rate of the secondary $i$ and is given by
\begin{equation}\label{eq:Qsecondaries}
Q_{i}=\sum_{j\neq i}n_j\frac{\sig_{j\ra i}}{m_p}c\rho_{\text{ISM}},
\end{equation}
where $\sig_{j\ra i}$ is the decayed spallation cross section of the parent nuclei $j$ into the secondary $i$ per ISM
nucleon. To see this, note that since the composition is the same at any given time or place, the ratio of the production
rates will be the same everywhere. For every particle B that is generated, $Q_{A}/Q_{B}$ particles of type A are
generated. As the particles propagate in the same manner, the density of A will be $Q_{A}/Q_{B}$ times that of B.

It is useful to write equation \eqref{eq:SecondaryRatioEquationNS} as
\begin{equation}\label{eq:GrammageNS}
n_i(\vep)=\frac{Q_i(\vep)}{\rho_{\text{ISM}}c} X_{\esc}(\vep/Z),
\end{equation}
where the grammage $X_{\text{esc}}$, defined by this equation, parameterizes the column density of target material
traversed by the CRs and is the same for all species. $Q_i(\vep)/(\rho_{\text{ISM}}c)$ is the local net generation of the
CRs per unit traversed mass, and is independent of the local ISM density. Note that $n_j$ are directly measured by CR
experiments and $\sig_j,\sig_{j\ra i}$ are measured in the laboratory. Thus, $X_{\text{esc}}$ can be directly extracted
from compositional measurements of CRs using Eqs. \eqref{eq:Qsecondaries} and \eqref{eq:GrammageNS}. Once $\Xesc$ is
determined, the density of any secondary can be computed using these equations along with the primary CR and
the cross section measurements.

\subsection{Stable secondaries that suffer spallation losses}\label{sec:SimpleSpal}
In reality, all measured spallation secondary nuclei suffer significant losses due to spallation.
Equations \eqref{eq:SecondaryRatioEquationNS}-\eqref{eq:GrammageNS} can be generalized to include spallation by replacing
the production rates $Q_i(\vep)/(\rho_{\text{ISM}}c)$ with net production rates $\tilde Q_i$ given by
\begin{equation}\label{eq:GenerationRate}
\tilde Q_i=\frac{Q_i}{\rho_{\text{ISM}} c}-\frac{n_i\sig_i}{m_p},
\end{equation}
where $\sig_i$ is the cross section for destruction of the CR per ISM nucleon and is approximately given by
$\sig_i\approx 40 A_i^{0.7}\mb$, with $A_i$ the atomic number of the particle i. This prescription results with
\begin{equation}\label{eq:SecondaryRatioEquation}
\frac{n_A}{n_B}=\frac{\tilde Q_{A}}{\tilde Q_{B}}
\end{equation}
and
\begin{equation}\label{eq:Grammage}
n_i(\vep)=\tilde Q_i(\vep) X_{\esc}(\vep/Z).
\end{equation}
To see this, note that the distribution of particles A that suffer spallation losses is equal to the distribution of
particles A' that do not suffer spallation if the production of A' is equal at any given time and place to the net
production of A.

Equations \eqref{eq:GenerationRate} and \eqref{eq:Grammage} can be written in a directly applicable form as:
\begin{equation}\label{eq:Grammage2}
n_i=\frac{X_{\esc}Q_i/(\rho_{\text{ISM}}c)}{1+\frac{\sig_i}{m_p}X_{\esc}}.
\end{equation}

As far as we know, Eq. \eqref{eq:Grammage2} is consistent with all data of CR composition. Above a few
$\GeV/\text{nuc}$, the value of $X_{\text{esc}}$ is measured to be
\begin{equation}\label{eq:X}
X_{\text{esc}}\approx 8.7 \left(\frac{\vep}{10Z \GeV}\right)^{-0.5}\gr\cm^{-2},
\end{equation}
with different fits varying by $\sim 30\%$ in the range $10\GeV<\vep/Z\lesssim100\GeV$
\citep[e.g.][]{Engelman90,Jones01,Webber03}. There are indications that the power law behavior of $\Xesc$ continues to
hundreds of GeV \citep{Binns88,Ahn08}, see however \citep{Zatespin09}. Henceforth, we assume that the grammage
parametrization given in Eq.\eqref{eq:X} holds up to $\vep/Z\sim300\GeV$.

For strong spallation losses, $\sig\gg m_p/\Xesc$, the density of secondaries given by Eq. \eqref{eq:Grammage2}
approaches a value that is independent of $\Xesc$,
\begin{equation}\label{eq:ninfinity}
n_{i,\infty}=\frac{Q_im_p}{\rho_{\text{ISM}}\sig_ic}.
\end{equation}
The strong suppression due to spallation of the heavy secondaries (Sc, V, Ti), results in a small deviation of their
densities from the limit of infinite grammage
\begin{equation}
\frac{n_{i,\infty}-n_i}{n_i}=\frac{m_p}{\Xesc\sig_i}\approx 0.3
\left(\frac{A_i}{50}\right)^{-0.7}\left(\frac{\vep}{10\GeV Z}\right)^{0.5}.
\end{equation}
Thus, the measurements of these sub-iron elements are useful for determining $\Xesc$ only at high energies
$\vep/Z\gtrsim 100\GeV$.

What makes equation \eqref{eq:SecondaryRatioEquation}-\eqref{eq:Grammage2} non trivial, is the fact that the loss term
has to be included in the expression for the net generation rate, Eq.~\eqref{eq:GenerationRate}. As the net generation
rate of two particle species A and B is affected by their own density, their relative abundance is required to be uniform
in order that $\tilde Q_A/\tilde Q_B$ be uniform. Thus the validity of equations
\eqref{eq:SecondaryRatioEquation}-\eqref{eq:Grammage2} suggests that the relative abundance of the secondaries themselves
is uniform.

Perhaps, the simplest propagation model in which the above conditions are realized is the homogenous Leaky Box Model
(LBM, see \sref{sec:OtherModels}). Obviously this model satisfies the conditions \eqref{eq:Conditions} and thus equations
\eqref{eq:SecondaryRatioEquation}-\eqref{eq:Grammage2} are guaranteed to hold. It is also known that these equations are
satisfied for disc-halo diffusion models in which the radial extent of the cosmic ray halo is much larger than the scale
height, which is in turn much larger than the width of the gas disc \citep[e.g.][]{Ginzburg76,Schlickeiser85}. In this
case again, conditions \eqref{eq:Conditions} are trivially satisfied. Expressions for $\Xesc$ in these models are given
in \sref{sec:OtherModels} and \sref{sec:models}.

In our view, Equations \eqref{eq:SecondaryRatioEquation} - \eqref{eq:Grammage2} are natural relations, that are
expected for a wide range of models, that satisfy conditions \eqref{eq:Conditions} and that were empirically validated.
Diffusion models, that have a large halo and thin disc (virtually all currently used models) or leaky box models are
particular models that satisfy these conditions. In fact, these equations will hold in any 1-D model in which the gas is
concentrated in a thin disc, assuming that the transport of particles (diffusive or otherwise) depends on their rigidity
only and that their energies do not change.

The good news are that these equations allow us to obtain robust predictions for secondary particles like anti-protons and
positrons (with the latter requiring more care due to energy losses, see \sref{sec:simple} and \sref{sec:losses}). The
bad news are that the measurements of stable secondaries, which are probably the single most important type of
measurements for quantitative research of CR propagation, carry little information regarding the precise form in which CRs
propagate.

We conclude this section with a comment about the application of these equations to primary CRs. Unlike secondaries, the
source function of primary particles is not known. In fact, equation \eqref{eq:Grammage2} is used to deduce the averaged
source spectrum and in particular the total required energy output of CRs. The fact that the resulting source spectrum,
when using the grammage deduced from the secondary measurements, is approximately the same for the different elements
\citep[e.g.][]{Engelman90} suggests that the same equations are applicable to the primary CRs as well. This in turn
suggests that the propagation of CRs averages the generation spectrum over distances larger than the inhomogeneities of
the primary sources. We note that the application of these equations to primaries is somewhat less substantiated
theoretically and observationally than for the secondaries.

\section{Application to secondary anti-protons and positrons}\label{sec:simple}
In this section, we estimate the expected flux of positrons and anti-protons using the measured CR traversed grammage
discussed in \sref{sec:Grammage}. We first discuss the local production rates of positrons and anti-protons in
\sref{sec:Prod}. We then write down a model independent expected $\bar p$ flux in \sref{sec:pbar} and show that it agrees
with observations. An upper limit for the secondary $e^+$ flux, obtained by neglecting energy losses, is given in
\sref{sec:positrons} and compared to observations. The $e^+$ energy losses are addressed in \sref{sec:losses}.

We note that for positrons and anti-protons the second of conditions \eqref{eq:Conditions} is not satisfied, as these
particles are generated at rigidities lower by a factor of $\sim 10$ compared to their progenitors. The fluxes of
positrons and anti-protons are therefore more sensitive to spectral variations of the CRs in the Galaxy compared to
products of spallation of nuclei.

\subsection{Production of anti-protons and positrons}\label{sec:Prod}
The rate of production of positrons and anti-protons depends on the flux of primary CRs, the ISM composition and nuclear
cross section data. For $\vep\gtrsim$ few GeV, the cross section dependence on energy is essentially dictated by the
cross section for pp collisions, while the presence of heavy target and projectile nuclei can be approximated via an
energy independent scaling factor, denoted here by $\xi_{S,A>1}$, of order unity.

Concerning the production of anti-protons, we adopt the cross section parametrization of \citet{Tan83b}. Given a
measurement of the primary proton flux $J_p$, the $\bar p$ production rate per unit ISM particle mass is given by
\begin{equation}\label{qspal}Q_{\bar p}(\vep)=2\xi_{\bar p,A>1}4\pi\int_{\vep_{\bar p}}^\infty
d\vep_pJ_p(\vep_p)\left(\frac{d\sigma_{\bar p}(\vep_p,\vep)}{d\vep_p}\right),\end{equation}
where the factor of 2 accounts for the decay of antineutrons produced in the same interactions.

For positrons, the production rate is given by a formula similar to~(\ref{qspal}), using the cross section for final
state positrons resulting from the decay of charged mesons. For the charged meson cross section, we again adopt the
parametrization of \citet{Tan83b}. The subsequent $e^+$ yield is calculated using standard electroweak theory. The $e^+$
yield we find agrees with the results of \citet{Delahaye08} to $\sim 10\%$ for the same parametrization.

For a steeply declining, smooth primary proton spectrum, it is useful to parameterize the resulting local generation rate
of anti-protons and of positrons per ISM mass by the following equation:
\begin{equation}\label{eq:Q_S}
\vep Q_S(\vep)=\xi_{S,A>1}(\vep)C_{S,pp}(\vep) 4\pi (10\vep) J_p(10\vep)\frac{\sig_{pp,0}}{m_p},
\end{equation}
where $S=\bar p, e^+$ stands for anti-protons and positrons respectively, $\sig_{pp,0}\equiv 30\mb$ is a cross section
normalization chosen to be approximately the inelastic cross section for pp interactions at the energy range
$10\GeV<\vep<300\GeV$ \citep{Tan83a} and $C_{S,pp}(\vep)$ is a dimensionless coefficient that weakly depends on the
primary spectrum.

\begin{figure}\hspace{-1.5cm}
\epsscale{1.4} \plotone{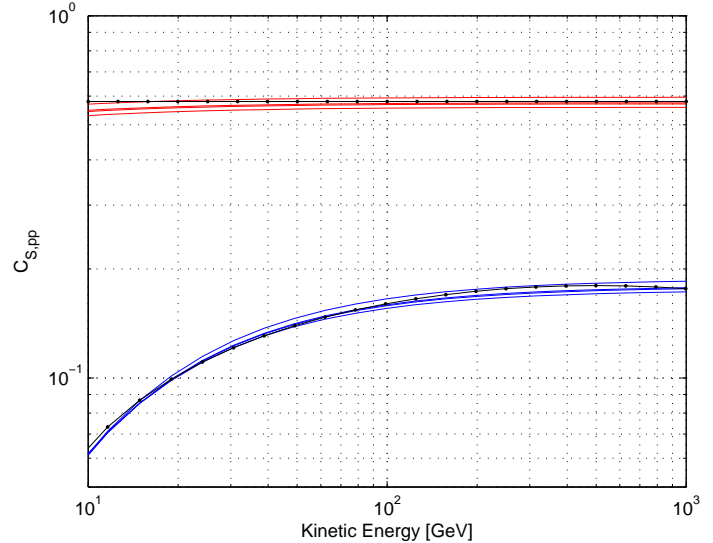}  \caption{Production coefficients $C_{S,pp}(\vep)$ [Eq. \eqref{eq:Q_S}] of anti-protons
and positrons. The lower blue curves are the production coefficients of anti-protons for proton spectra $\gamma=2.6-2.9$,
and the upper red curves are the production coefficients of positrons for the same values of $\gamma$. The black dotted
line is the approximation given in Eqs. \eqref{eq:Cpbar} and \eqref{eq:Cpos}.} \label{fig:CSpp}
\end{figure}

The values of $C_{\bar p,pp}$ and $C_{e^+,pp}$ are shown in figure \ref{fig:CSpp} for a power law proton flux $J_p\propto
\vep^{-\gamma}$ with $2.6<\gamma<2.9$.
For this range of values for $\gamma$, the approximation
\begin{equation}\label{eq:Cpbar}
C_{\bar p,pp}=0.18-0.04\log^2_{10}(\vep/500\GeV),
\end{equation}
and
\begin{equation}\label{eq:Cpos}
C_{e^+,pp}=0.58
\end{equation}
at the $\bar p$ ($e^+$) kinetic energies $10\GeV<\vep<\TeV$, is accurate to better than $10\%$ (approximation shown in
the figure).

For completeness, we also give the approximate value of $C_{e^+,pp}$ corresponding to the parametrization of
\citet{Kamae06}
\begin{equation}\label{eq:Cpos}
C_{e^+,pp}=0.31+0.15\log_{10}(\vep/100\GeV)\Te(\vep-100\GeV),
\end{equation}
valid in the range $2.6<\gamma<2.9$ and $10\GeV<\vep<1\TeV$ to an accuracy of 10\%.

\subsection{Anti-Protons}\label{sec:pbar}
Using equations \eqref{eq:Grammage2} and \eqref{eq:Q_S}, and assuming a power-law proton spectrum $J_p\propto
\vep^{-\gamma}$, the expected ratio of anti-protons to protons is given by
\begin{equation}\label{eq:PbarOverP}
\frac{J_{\bar p}}{J_p}=10^{-\gamma+1}\xi_{\bar p,A>1}C_{\bar
p,pp}(\vep)\frac{\sig_{pp,inel,0}}{m_p}X_{\esc}\frac{1}{1+\frac{\sig_{\bar p}}{m_p}X_{\esc}}.
\end{equation}

\begin{figure}\hspace{-2cm}
\epsscale{1.4} \plotone{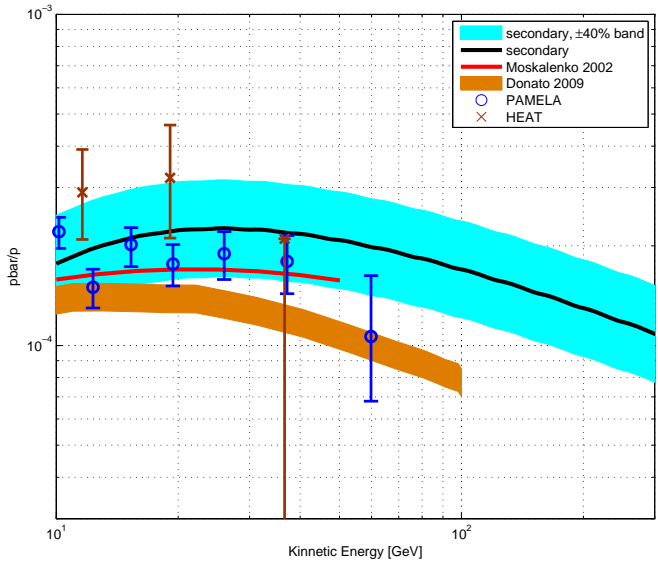} \caption{Anti-proton to proton flux ratio. The blue and dark brown error-bars are
the $J_{\bar p}/J_{p}$ ratios measured by PAMELA \citep{Adriani09b} and HEAT \citep{Beach01}. The black line is the
expected ratio using equation \eqref{eq:PbarOverP}, with the cyan colored region denoting a 40$\%$ \citep{Simon98}
uncertainty band. The brown band depicts the results of the diffusion model of \citet{Donato09}. \label{fig:PbarOverP}}
\end{figure}

Using the measured value of $\Xesc$ from Eq.\eqref{eq:X}, setting $\xi_{\bar p,A>1}=1.2$ ~\citep{Gaisser92,Simon98}, we find
\begin{align}
\frac{J_{\bar p}}{J_p}\approx&3.6\times 10^{-4}\left(\frac{C_{\bar p
,pp}(\vep)}{0.1}\right)\left(\frac{\vep}{10\GeV}\right)^{-0.5}\cr &\times\left(1+0.16\left(\frac{\sig_{\bar
p}}{30\mb}\right)\left(\frac{\vep}{10\GeV}\right)^{-0.5}\right)^{-1}.
\end{align}
In our calculation we adopt $\sig_{\bar p}$ from \citet{Tan83a}, where $\sig_{\bar p}\approx 30\mb$ holds to an accuracy
of about $20\%$ in the range $10\GeV<\vep<100\GeV$. This result is compared to experiments in figure \ref{fig:PbarOverP}.
As can be seen, the results are in good agreement with recent measurements \citep{Adriani09b,Beach01}.

Similar calculations, up to differences in the cross section and grammage parameterizations, were made in
\citep{Gaisser92,Simon98}. In figure \ref{fig:PbarOverP}, the results of two detailed diffusion models \citep{Donato09,
Moskalenko02} are plotted. The model of \citep{Donato09} assumes a primary proton spectral index of $\gamma=2.84$, cross
sections ate taken from \citet{Bringmann07} with the p-p cross sections based on \citet{Tan83a} \citep[which are slightly
smaller than the parametrization by][that we use]{Tan83b}, and uses an energy dependent diffusion coefficient
$D(\vep)\propto \vep^{0.7}$. The model of \citep{Moskalenko02} assumes a primary proton spectral index of $\gamma\approx
2.75$, $pp$ cross sections based on \citet{Tan83b}, and uses an energy dependent diffusion coefficient $D(\vep)\propto
\vep^{0.6}$. The results of the different computations agree to within a factor of $\sim2$.

\subsection{Positrons}\label{sec:positrons}
Using equations \eqref{eq:fDef},\eqref{eq:Grammage2} and \eqref{eq:Q_S} and assuming a power-law proton spectrum
$J_p\propto \vep^{-\gamma}$, the expected ratio of positrons to protons is given by
\begin{equation}\label{eq:Jpos0}
\frac{J_{e^+}}{J_p}=f_{s,e^+}10^{-\gamma+1}\xi_{e^+,A>1}C_{e^+,pp}(\vep)\frac{\sig_{pp,inel,0}}{m_p} X_{\esc}.
\end{equation}
Using the measured value of $\Xesc$ from Eq.\eqref{eq:X}, setting $\xi_{e^+,A>1}\approx 1$ \citep{Moskalenko98}, we find
\begin{equation}
\frac{J_{e^+}}{J_p}\approx 1.7\times 10^{-3}f_{s,e^+}\left(\frac{C_{e^{+}
,pp}(\vep)}{0.6}\right)\left(\frac{\vep}{10\GeV}\right)^{-0.5}.
\end{equation}
Adopting a high energy Inter Stellar (IS) proton flux \citep{Moskalenko02},
\begin{equation}
J_p=1.6\times 10^4 (\vep/\GeV)^{-2.75} \GeV^{-1}\m^{-2}\sec^{-1}\sr^{-1},
\end{equation}
we have
\begin{align}\label{eq:Jpos}
& J_{e^+}\approx  4.7\times 10^{-2}f_{s,e^+}\left(\frac{C_{e^{+}
,pp}(\vep)}{0.6}\right)\left(\frac{\vep}{10\GeV}\right)^{-3.25}\cr &\GeV^{-1}\m^{-2}\sec^{-1}\sr^{-1}.
\end{align}

\begin{figure}[t]\hspace{-1.5cm} \epsscale{1.4} \plotone{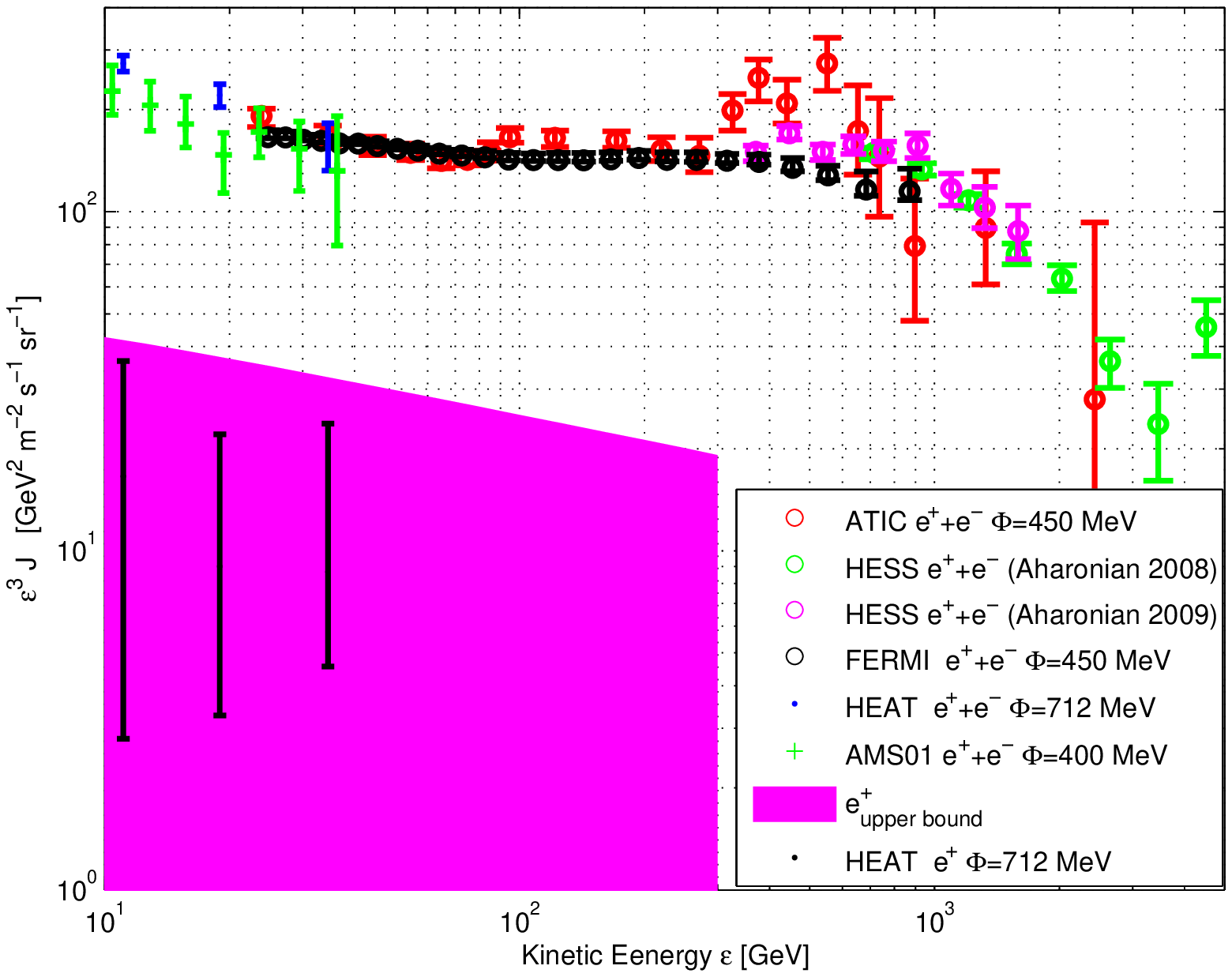}
\caption{Interstellar $e^+$ flux measured by HEAT \citep{DuVernois01} compared with the theoretical upper bound, obtained
from Eq.\eqref{eq:Jpos} with $f_{s,e^+}=1$ for $\vep<300\GeV$. The allowed value of the $e^+$ flux, corresponding to
$f_{s,e^+}<1$ is shown as a shaded region. Also plotted is the $e^+ + e^-$ flux as measured by HEAT \citep{DuVernois01},
FERMI \citep{Abdo09}, ATIC \citep{Chang08} and AMS01\citep{Alcaraz00}. Fluxes are demodulated using the indicated values
of the solar modulation potential $\Phi$.} \label{fig:epose}
\end{figure}

An upper limit for the $e^+$ flux, obtained by setting $f_{s,e^+}=1$ in equation \eqref{eq:Jpos}, is shown in figure
\ref{fig:epose}
as a shaded region along with $e^+$ and $e^+ + e^-$ flux measurements. As can be seen, the $e^+$ data are compatible with
a secondary origin. The data implies a suppression factor of order $f_{s,e^+}(\vep)\sim 1/\text{few}$ (see discussion of
$e^+$ energy losses in \sref{sec:losses}).

The positron fraction was measured by several experiments including the recent PAMELA experiment. Since the $e^-$ flux is
poorly understood theoretically, we do not have a model independent prediction for the positron fraction. In order to
compare our predictions for the $e^+$ flux to the positron fraction measurements we use $e^+ + e^-$ measurements.

\begin{figure}[t]\hspace{-2cm} \epsscale{1.4} \plotone{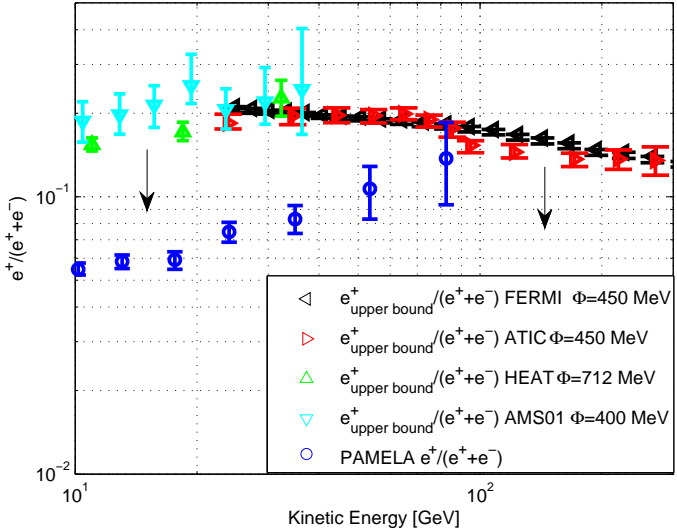}
\caption{Interstellar positron fraction as measured by PAMELA \citep{Adriani09a} compared with an upper limit for the
positron fraction. The upper limit is obtained by dividing the theoretical $e^+$ flux upper bound with $e^+ + e^-$ fluxes
measured by HEAT \citep{DuVernois01}, FERMI \citep{Abdo09}, ATIC \citep{Chang08} and AMS01 \citep{Alcaraz00}.}
\label{fig:PAMELA}
\end{figure}

The claimed PAMELA excess extends from sub $10\GeV$ energies with high statistics to $100\GeV$ with low statistics. The
$e^+ + e^-$ spectrum is measured by different experiments in different energy intervals within this range. Thus, it is
useful to compare the PAMELA measurement with the various $e^+ + e^-$ measurements and the theoretical expectation
\eqref{eq:Jpos} using a single plot. This is done in figure \ref{fig:PAMELA},
where we divide our $e^+$ flux upper bound derived from \eqref{eq:Jpos} by the measured $e^+ + e^-$ flux and compare the
results to the PAMELA positron fraction data.

\begin{figure}\hspace{-2cm}
\epsscale{1.4} \plotone{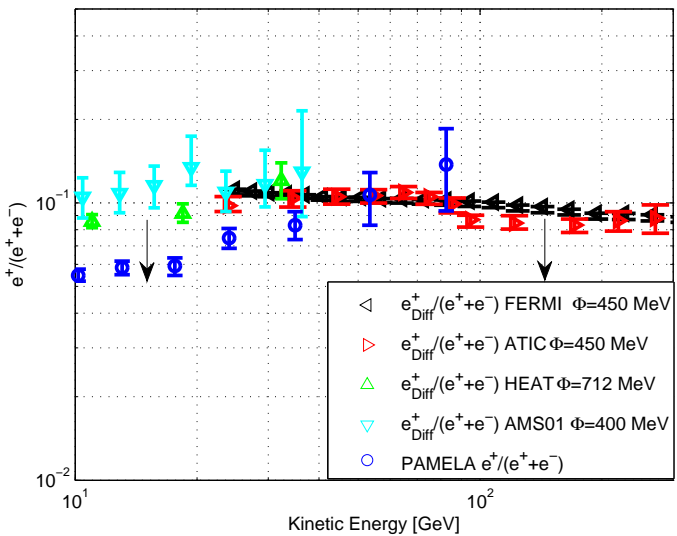} \caption{As figure \ref{fig:PAMELAKamae}, using the cross section
parametrization of \citep{Kamae06}.}\label{fig:PAMELAKamae}
\end{figure}

For completeness, in figure \ref{fig:PAMELAKamae} we reproduce the same calculation leading to figure \ref{fig:PAMELA},
using the $e^+$ production cross section parametrization from \citep{Kamae06}. This recent parametrization results with a
$e^+$ yield that is lower by about a factor of two compared to the parametrization from \citep{Tan83b}, used throughout
this paper.

As can be seen, the PAMELA and $e^+ + e^-$ measurements are consistent with a secondary origin for the positrons
throughout the measured energy range, with a reasonable suppression factor  $f_{s,e^+}^{-1}(\vep)\sim 1-4$ (see
discussion of $e^+$ energy losses in \sref{sec:losses}).

As can be seen in figures \ref{fig:PAMELA} and \ref{fig:PAMELAKamae}, the positron fraction cannot exceed
$e^+/(e^++e^-)\approx 0.2$ in the energy range $100\GeV<\vep<300\GeV$. This is based on the measured $e^++e^-$ flux by
FERMI and ATIC, the measured CR grammage Eq. \eqref{eq:X}, limited by currently available data to hundreds of GeV, and
the fact that $f_{s,e^+}<1$.

In the energy interval $\vep\gtrsim 30\GeV$, there is an indication from the measurements that $f_{s,e^+}(\vep)$ is
rising with energy. Since $f_{s,e^+}<1$, this trend cannot continue indefinitely. If $f_{s,e^+}$ is rising with energy, a
break in the rising pattern must occur at $\vep\lesssim 100\GeV$. If primary electrons are affected by the energy losses
which affect the secondary positrons, a feature in the $e^+ + e^-$ flux is expected. Note that at higher energies the
statistical and systematic errors in the positron fraction are large, and thus the rising trend of $f_{s,e^+}$ is not
implied by the observations with high significance.

\subsection{Positrons/Anti-Protons}
Both positrons and anti-protons at a given energy $\vep$ are produced by the interactions of the same primary CRs (mainly
protons) at roughly the same primary energies (roughly $10\vep$). Hence, the $e^+$ to $\bar p$ flux ratio is particularly
insensitive to the compositional and spectral variations of the CRs.

\begin{figure}
\epsscale{1.3} \plotone{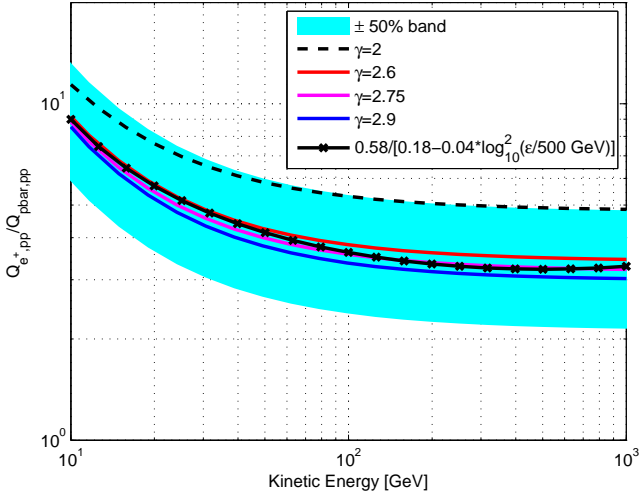} \caption{$e^+$ to $\bar p$ production ratio due to pp interactions, which is
also the flux ratio upper limit given by equation \eqref{eq:epbarineq}. The different smooth, colored lines are for
different values of the proton spectral index $\gamma$, in the range $2.6<\gamma<2.9$. The crossed black line is the
expected ratio using equations \eqref{eq:Cpbar} and \eqref{eq:Cpos}. The dashed black line corresponds to an extreme
value $\gamma=2$. The cyan colored band depicts 50$\%$ modulation over the central value for $\gamma=2.75$, to be
regarded as a conservative uncertainty estimate. \label{fig:epbar}}
\end{figure}

 The ratio of positrons to anti-protons is given by:
\begin{equation}\label{eq:epbar}
\frac{J_{e^+}}{J_{\bar p}}=\left(\frac{\xi_{e^+,A>1}}{\xi_{\bar p,A>1}}\right)\left(\frac{1}{1+\frac{\sig_{\bar
p}}{m_p}X_{\esc}}\right)f_{s,e^+}\frac{C_{e^+,pp}(\vep)}{C_{\bar p,pp}(\vep)}.
\end{equation}
Both $\xi_{e^+,A>1}$ and $\xi_{\bar p,A>1}$ are not very different from $1$. Measurements suggest that
$\xi_{e^+,A>1}<\xi_{\bar p,A>1}$ \citep{Adler06}. The $\bar p$ losses are not significant above $10\GeV$ (about $15\%$ at
$10\GeV$ and droping with energy). Quite generally we thus have:
\begin{equation}\label{eq:epbarineq}
\frac{J_{e^+}}{J_{\bar p}}\lesssim \frac{C_{e^+,pp}(\vep)}{C_{\bar p,pp}(\vep)}=\frac{Q_{e^+,pp}}{Q_{\bar p,pp}},
\end{equation}
where $Q_{S,pp}$ is the generation rate of specie $S$ in pp interactions. The $e^+$ to $\bar p$ branching ratio,
$Q_{e^+,pp}/Q_{\bar p,pp}$, is calculated based on particle physics and accelerator measurements (see section
\sref{sec:Prod}) and is shown in figure \ref{fig:epbar}. As seen in the figure, this ratio weakly depends on energy above
a few tens of $\GeV$ and is given roughly by $Q_{e^+,pp}/Q_{\bar p,pp}\sim3$. Furthermore, this ratio depends weakly on
the primary proton spectrum. To illustrate this fact, we also plot in figure \ref{fig:epbar} the ratio resulting from an
extreme value of the proton spectral index, $\gamma=2$. We find that $Q_{e^+,pp}/Q_{\bar p,pp}$ remains within $50\%$
from its value for $\gamma=2.75$.

The $e^+$ to $\bar p$ flux ratio is not given in any experimental report that we are aware of. In order to compare the
expected ratio, in the energy range $\vep\sim 10-100\GeV$, with the ratio inferred from the PAMELA measurements of
$e^+/(e^++e^-)$ and $\bar p/p$ combined with proton and $e^++e^-$ measurements, it is useful to parameterize the ratio
as:
\begin{align}\label{eq:postopbarMes}
&\frac{J_{e^+}}{J_{\bar p}}=2
\left(\frac{J_{e^+}/(J_{e^+}+J_{e^-})}{0.1}\right)\left(\frac{(J_{e^+}+J_{e^-})\vep^3}{150\GeV^2 \m^{-2}
\se^{-1}\sr^{-1}}\right)\cr &\left(\frac{J_p\vep^{2.75}}{1.6\times 10^{4}\GeV^{1.75} \m^{-2}
\se^{-1}\sr^{-1}}\right)^{-1}\left(\frac{J_{\bar p}/J_p}{2\times10^{-4}}\right)^{-1}\cr
&\left(\frac{\vep}{30\GeV}\right)^{-0.25}.\cr
\end{align}
All the factors in Eq. \eqref{eq:postopbarMes} are measured to be close to unity in the relevant energy range. Hence the
experimental results are consistent with  Eq.\eqref{eq:epbar} for $f_{s,e^+}^{-1}\sim2-3$. Obviously, a direct
measurement of the $e^+$ to $\bar p$ flux ratio would be much more useful.

\section{Propagation time scales and positron energy losses}\label{sec:losses}
The local density of secondaries that decay or suffer energy loss will be suppressed compared to the prediction of
Eq.\eqref{eq:Grammage} if their decay/loss time is short compared to their typical residence time. Positrons lose energy
by synchrotron and Inverse Compton (IC) processes with a cooling time
\begin{align}\label{eq:posLossICSyn}
&\taucool=\frac{\vep}{\dot\vep}=\frac{m_e^2c^4}{\frac43\vep\sig_T U_Tc}\approx 10^7\yr\cr &\times
\left(\frac{\vep}{30\GeV}\right)^{-1}\left(\frac{U_T}{1\eV\cm^{-3}}\right)^{-1},
\end{align}
where $U_T$ is the sum of the ambient photon and magnetic field energy densities. This time scale is comparable to the CR
propagation time scale at the energy range of interest, so some amount of cooling suppression of positrons is expected.
The radiative loss suppression depends on details of the CR propagation in the Galaxy, which is not understood
theoretically.

Measurements of the abundance of radioactively unstable CR nuclei with known decay times give valuable information on the
residence time scale of the CRs and hence on the suppression of electrons and positrons due to cooling. In this section
we use the results of such measurements to constrain the value of $f_{s,e^+}$ [see Eq. \eqref{eq:fDef}]. In
\sref{sec:lossesTheory} we explain how measurements of the suppression factor of radioactive nuclei can be used to
constrain the cooling suppression of electrons in a model insensitive way. In particular we address the complications
that arise due to the spallation of the radioactive nuclei and to the difference in the suppression mechanism (decay vs.
energy loss). In \sref{sec:lossesAp} we discuss the results of abundance measurements and give an order of magnitude
estimate of $f_{s,e^+}$. This estimate is shown to be consistent with the results of \sref{sec:simple}. A discussion of
$f_{s,e^+}$ in common models is given in \sref{sec:lossesmodels}.

\subsection{Radioactive decay vs radiative energy loss}\label{sec:lossesTheory}
As in the case of electrons and positrons, it is useful to separate the effect of decay and propagation for unstable nuclei. A complication that arises in the case of nuclei is the fact that in addition to
decay, the nuclei are destroyed by spallation. It is useful to separate the effect of losses due to decay from that of
spallation. This can be done in a model independent way by considering the net production of the particles instead of the
source production, and defining the decay suppression factor $f_{s,i}$ by
\begin{equation}\label{eq:Deffi}
n_i=f_{s,i}\tilde Q_i\Xesc=f_{s,i}\left[Q_i/(\rho_{\text{ISM}}c)-n_i\frac{\sig_i}{m_p}\right]\Xesc,
\end{equation}
where  Eqs. \eqref{eq:GenerationRate}  and \eqref{eq:Grammage} were used. This suppression factor is more useful for
obtaining the suppression factor of positrons compared to the more commonly used surviving fraction $\tilde f_{s,i}$
\citep[e.g.][]{Webber98}, defined as
\begin{equation}\label{eq:Detffi}
n_i=\tilde f_{s,i}n_{i,\text{no decay}},
\end{equation}
were $n_{i,\text{no decay}}$ is the density of this nuclei that would result if it did not suffer decay (but did suffer
spallation losses). Using Eq.\eqref{eq:Grammage2}, we find that $n_{i,\text{no decay}}$ is given by
\begin{equation}\label{eq:Detffi}
n_i=\tilde f_{s,i}n_{i,\text{no decay}}=\tilde f_{s,i}\frac{(Q_i/\rho_{\text{ISM}})\Xesc}{1+\frac{\sig_i}{m_p}\Xesc}.
\end{equation}
Substituting $n_i$ from Eq. \eqref{eq:Detffi} in Eq. \eqref{eq:Deffi} and extracting $f_{s,i}$, we find that the two
factors are related by:
\begin{equation}\label{eq:ftotf}
\frac{f_{s,i}}{\tilde f_{s,i}}=\left[1+\frac{\sig_i}{m_p}\Xesc(1-\tilde f_{s,i})\right]^{-1}.
\end{equation}

Next we demonstrate that under a wide range of conditions, the suppression factor $f_{s,i}(\vep/Z)$ of a nucleous that
suffers decay with a decay time $\tau_d$, is similar to the suppression factor $f_{s,e^+}(\vep)$ that positrons suffer
due to energy losses with a cooling time $\tau_c$, if $\tau_c\sim\tau_d$ at the same rigidity. The loss rates due to
decay and energy losses in a general transport equation (assuming that the distribution is close to isotropic) are given
by
\begin{equation}\label{eq:Declossi}
\pr_t n_{i,\text{decay}}=-\frac{n_i}{\tau_d}
\end{equation}
and
\begin{equation}\label{eq:poslossi}
\pr_t n_{e^+,\text{energy loss}}=\pr_\vep\left(\dot \vep n_{e^+}\right)
\end{equation}
respectively. For losses due to IC and synchrotron where $\dot\vep\propto\vep^2$, we can rewrite Eq.\eqref{eq:poslossi}
as
\begin{equation}\label{eq:poslossi2}
\pr_t n_{e^+,\text{energy loss}}=\left(\frac{\pr\log(\vep^2n_{e^+})}{\pr\log(\vep)}\right)\frac{n_{e^+}}{\tau_c}.
\end{equation}
Assuming that the $e^+$ spectrum is steeper than $n\propto \vep^{-2}$, the term in parentheses is negative and of order
unity. The energy loss term Eq.\eqref{eq:poslossi2} is similar to the decay term Eq. \eqref{eq:Declossi} with $\tau_d$
replaced by $\tau_c$, up to an order unity correction. In particular, for a spectrum $n\propto \vep^{-\gamma}$ this term
equals $\gamma-2$ and is unity for $\gamma=3$.

The similarity between cooling and decaying can be understood as follows. During a time interval $\tau_c(\vep)$, a $e^+$
with energy $\vep$ loses a significant part of its energy. Assuming that the $e^+$ spectrum is steeper than $n\propto
\vep^{-2}$, the contribution of such positrons to the lower energy flux is negligible and they are effectively lost.

Since the loss terms of positrons and unstable nuclei with $\tau_d\sim \tau_c$ are similar, we expect that their
suppression factors will be similar. In \ref{sec:lossesmodels} we show how this is realized within specific models.

\subsection{Measurements and constraints on $f_{s,e^+}$}\label{sec:lossesAp}
Perhaps the single most important unstable isotope which is used for propagation time scale estimates is $^{10}$Be. This
is due to the fact that (a) it is a pure secondary and (b) it has a lifetime $\tau_{\rm rest}\approx 2.2\times
10^{6}\yr$, which turns out to be of the same order of magnitude as the escape time. Other commonly used unstable nuclei
include $^{14}$C, $^{26}$Al, $^{36}$Cl and $^{54}$Mn \citep[see e.g.][]{Strong07}.

Direct measurements of the abundance of these isotopes exist only at sub-GeV/nuc energies. In fact, measurements with
high statistics are limited to a narrow energy range around $\sim 100$ MeV/nuc, where solar modulation and other low energy complications are important. Nevertheless, these low energy data are often used to
calibrate model parameters, which are later used to predict fluxes at all energies
\citep[e.g.][and references within.]{Berezinskii90,Moskalenko98,Strong01}.

An indirect measurement of the surviving fraction can be made by using the high energy, charge (as opposed to isotopic)
composition of CRs. For example, by measuring the charge ratio Be$/$B, and comparing it to the expected ratio that would
result if $^{10}$Be decayed completely or didn't decay at all, the suppression of $^{10}$Be can be extracted. Using this
method, the surviving fractions at $\vep \lesssim 15\GeV/$nuc (rigidity of $\vep/eZ\lesssim 30\GV$), were measured for a
few unstable isotopes \citep{Webber98}. In particular, the surviving fractions of $^{10}Be$ at energies $\vep\approx
4(16)\GeV/$nuc [rigidities of $\vep/eZ=10(40)\GV$]  were measured to be $\tilde f_{s,^{10}\text{Be}}\approx
0.4(0.5)\pm0.1$. Using Eqs.~\eqref{eq:X} and \eqref{eq:ftotf}, and adopting $\sig_{^{10}\text{Be}}\approx 200\mb$, the
suppression factors are found to be $f_{s,^{10}\text{Be}}\approx 0.3(0.4)$. The decay time of $^{10}$Be in the observer
frame at these two energies is $\tau_d\approx 0.88(2.2)\times 10^{7}\yr$. These decay times are similar to the cooling
times of positrons at the same rigidities, $\tau_c\approx 3.1(0.78)\times 10^{7}\yr (U_T/\eV)^{-1}$, as given by
Eq.\eqref{eq:posLossICSyn}.

Similar measurements of $^{26}$Al, $^{36}$Cl and $^{54}$Mn, that were derived based on elemental abundances of Al,Cl and
Mn, give similar results \citep{Webber98}. The analysis using these isotopes suffers additional uncertainties since Al,
Cl and Mn are not purely secondary.

Note that these measurements are particularly sensitive to uncertainties in the spallation cross sections. For example,
the Be$/$B ratio changes by about $20\%$ only between no $^{10}$Be decay vs complete decay. An accuracy of the cross
sections considerably better than $10\%$ is required. The resulting systematic uncertainties are hard to estimate. The
fact that the different measurements seem to be consistent with each other give some support to the derived values of
$f_{s,e^+}$.

Assuming that these measurements are valid, we conclude that at energies $\vep\sim 20\GeV$ the suppression factor of the
positrons is $f^{-1}_{s,e^+}\sim 3$, consistent with the results of the $e^+$ measurements presented in section
\sref{sec:simple}.

\section{Comparison to detailed models}\label{sec:OtherModels}
In this section we describe common models that are used in the literature. In \sref{sec:OtherModelsGrammage} we describe
the leaky box and disk-halo diffusion models and write down simple expressions for the grammage in these models. We show
that detailed models, with parameters chosen to fit compositional data, have a grammage that agrees with Eq.\eqref{eq:X}
at high energies. This is done by analyzing the results of the parameter scan made given \citet{Maurin01}. In
\sref{sec:lossesmodels} we write down expressions for the loss suppression factors. We show that our analytic estimates,
using the model's suppression factors, reproduce the results of the detailed calculations of the $e^+$ flux carried out
by \citet{Moskalenko98} and \citet{Delahaye08} to within a factor of $\sim 2$. As commonly claimed \citep{Adriani09a},
these results contradict PAMELA measurements. We highlight the assumptions built into these models, that lead to this
disagreement, and argue that they are not supported by observations.

More details of specific models can be found in \sref{sec:models} and in the extensive literature  \citep[for reviews
see][]{Ginzburg76,Berezinskii90,Strong07}.

\subsection{Grammage in common models}\label{sec:OtherModelsGrammage}
The CR grammage defined by equations \eqref{eq:Grammage}-\eqref{eq:Grammage2} is useful as long as the conditions for the
validity of these equations are fulfilled. As these equations are established observationally, they must be approximately
satisfied by any model that is consistent with compositional CR measurements. These equations will not be valid for a
general choice of model parameters, in which case the grammage is not well defined. For concreteness, we define the
grammage in a general model using Eq. \eqref{eq:GrammageNS}, for particles with no losses, even if the model does not
satisfy equations \eqref{eq:Grammage}-\eqref{eq:Grammage2} exactly. When necessary we specify the distribution of sources
assumed for this purpose.

In our calculations, we neglect changes in the particle energy resulting from assumed reacceleration or convection.
Radiative energy losses of positrons are treated in \sref{sec:lossesmodels}.

\subsubsection{Leaky Box Model}
In this model, the ISM and CRs are distributed homogenously in the galactic disc ('box'), from which CRs constantly
'leak' \citep{Ginzburg76}. This model has two free parameters: an effective averaged ISM density $\rho_{\text{LBM}}$ and
a rigidity dependent CR escape time $\tau_{\esc}(\vep/Z)$. The steady state equation for stable particles that do not
suffer spallation is
\begin{equation}\label{eq:LeakyBox}
Q_i=\frac{n_i}{\tau_{\esc}}.
\end{equation}
Equation \eqref{eq:LeakyBox} is equivalent to equation \eqref{eq:GrammageNS}, with
\begin{equation}\label{eq:XLBM}
\Xesc=\rho_{\text{LBM}}\tau_{\esc}c.
\end{equation}
As this model has a uniform composition for all particles, primary and secondary, the effect of spallation is exactly
described by equations \eqref{eq:SecondaryRatioEquation}-\eqref{eq:Grammage2}.

\subsubsection{Diffusion models}
It is commonly assumed that the propagation of the CRs is described by a diffusion equation. Models with varying
assumptions regarding the spatial distribution of the diffusion coefficient, sources and boundary conditions of the
propagation volume were studied. In many models, the CRs diffuse in a CR halo that surrounds the Galaxy, typically a
cylinder of radius $R\gtrsim 15\kpc$ and semi-height $L\sim \text{few}\kpc$, with a rigidity dependent diffusion
coefficient, while the ISM gas and the CR sources are distributed in the galactic gaseous disc with semi-height $h\sim
100 pc$. The CR density for all CR species is taken to vanish on the boundaries of the halo. Typically, the diffusion
coefficient is chosen to be spatially homogenous and depends on rigidity as $D=D_0(\vep/10ZGeV)^\de$, with
$D_0\sim10^{28}cm^2s^{-1}$ and $\delta\approx 0.5$.

In many examples it is assumed that $R\gg L\gg h$, in which case the radial boundary can be taken as infinity and the
disc as infinitely thin. In this case, we are left with a one dimensional model where the distribution of source
functions of all CRs is the same, namely a delta function in z. In this limit, which serves as a good approximation for
most purposes, the conditions \eqref{eq:Conditions} are exactly satisfied. The density of a particle $i$ which suffers no
losses at a distance $z$ from the disc is given by
\begin{equation}\label{eq:DiffDH}
n(z,\vep)=\frac{q_{i}L}{2D}(1-z/L),
\end{equation}
where $q_{i}$ is the generation rate per unit disc area. Equation \eqref{eq:DiffDH} is equivalent to equation
\eqref{eq:GrammageNS} when applied to the local position $z=0$ with the CR grammage given by
\begin{equation}\label{eq:XDiffDH}
\Xesc=\Xdisc Lc/(2D),
\end{equation}
where $\Xdisc\sim 10^{-3}\gr\cm^{-3}$ is the grammage of the gaseous disc and the generation rate per ISM mass is given
by $Q_i/\rho_{\text{ISM}}=q_i/\Xdisc$. The effect of spallation can be accurately taken into account by equations
\eqref{eq:SecondaryRatioEquation}-\eqref{eq:Grammage2}.

To illustrate the robustness of eq. \eqref{eq:Grammage}, as well as to demonstrate the calculation of the CR grammage in
a more complicated set-up, we next consider the detailed study of diffusion models made by \citet{Maurin01}. The results
of this study are widely used in recent publications. In this study, a $\chi^2$ analysis over $B/C$ data was made for a
wide range of values of the different parameters of the model. The model assumes a cylindrical geometry with radius
$R=20\kpc$, halo semi-height $1\kpc<L<15\kpc$, a diffusion coefficient given by
\begin{equation}\label{eq:Maurin01D}
D(\vep)=K_0\bt(pc/Z\GeV)^{\delta},
\end{equation}
a uniform convective wind with velocity $V_c$, and reacceleration parameterized by an Alfven velocity $V_A$.

\begin{figure}\hspace{-1cm}
\epsscale{1.3} \plotone{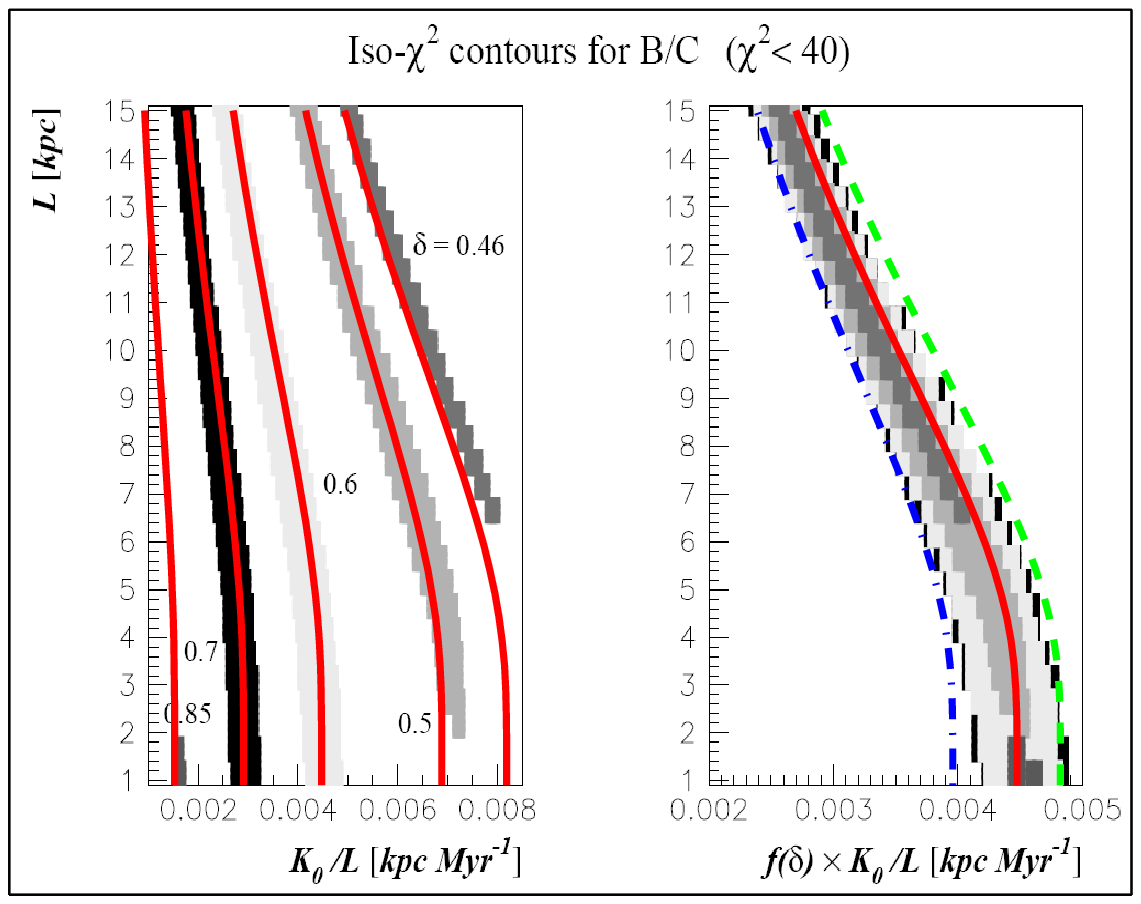}  \caption{Figure 7 from \citep{Maurin01X}. The filled gray areas on the left
panel display contour levels of $\chi^2<40$ for the indicated values of the diffusion power-law index $\delta$. The same
contour plots are shown on the right panel with the values of $K_0/L$ scaled by $f(\de)$ (see also text and figure
\ref{fig:MaurinFOfDelta}). Plotted on top of the figure are curves with a given grammage at the rigidity $\vep/Z=75 \GeV$
($\vep=35\GeV/nuc$), corresponding to the highest energy bin in the B/C data (see text). The red solid lines, green
dashed line and blue dash-doted line correspond to $\Xesc(75\GeV)=3.45\gr\cm^{-2}$, $\Xesc(75\GeV)=3.2\gr\cm^{-2}$ and
$\Xesc(75\GeV)=3.9\gr\cm^{-2}$ respectively.}\label{fig:Mourin01Fig7}
\end{figure}

\begin{figure}\hspace{-1.3cm}
\epsscale{1.2} \plotone{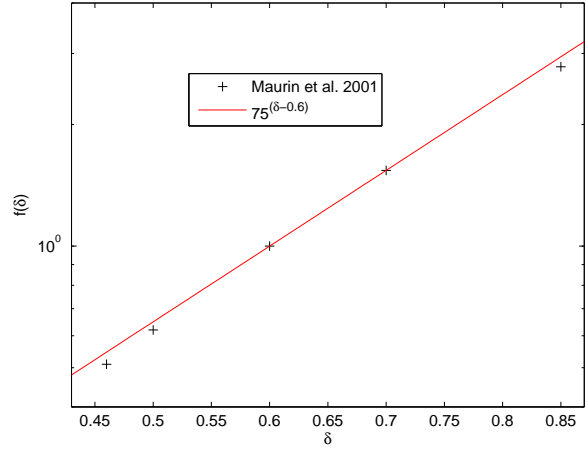}  \caption{The $K_0/L$ scaling function $f(\delta)$ \citep[][black pluses, see
also text and figure \ref{fig:Mourin01Fig7}]{Maurin01}. $f(\delta)$ is normalized to 1 at $\delta=0.6$.
$f(\de)=75^{\delta-6}$ (solid red line) is required in order that the grammage at the rigidity $\vep/Z=75 \GeV$
($\vep=35\GeV$/nuc), corresponding to the highest energy bin in the B/C data remain unchanged.}\label{fig:MaurinFOfDelta}
\end{figure}

The parameters $K_0, L$ and $\delta$ were found to be highly correlated. The low $\chi^2$ regions were found to be
described by the following relation
\begin{equation}\label{eq:Maurin01Corr}
K_0/L=\tilde g(L)/f(\de),
\end{equation}
and the numerical values of $f(\de)$ corresponding to several values of $\de$ were specified. This result can be seen in
figure \ref{fig:Mourin01Fig7} [taken from \citep{Maurin01X}]. The specified values of $f(\de)$ are shown in figure
\ref{fig:MaurinFOfDelta} as black crosses.

In terms of the analysis presented in the current paper, Eq.~\eqref{eq:Maurin01Corr} is explained as follows. The
reported correlations simply reflect the fact that, in order to fit the measured $B/C$ values, the model's grammage has
to agree with the measured result \eqref{eq:X}. More precisely the grammage constraint must hold at high energies, where
the energy changing effects of reacceleration and convection, that are not taken into account in
\eqref{eq:SecondaryRatioEquation}-\eqref{eq:Grammage2}, are negligible. In fact, the correlation \eqref{eq:Maurin01Corr}
is determined by the requirement that the value of the grammage at the highest energy point in the $B/C$ measurements,
$\vep=35\GeV$/nuc or $pc/Z\approx\vep/Z\approx 75\GeV$, be constant and approximately equal to the value given by
\eqref{eq:X}.

To illustrate this statement, we superimpose onto figure \ref{fig:Mourin01Fig7} contours of constant $X_{esc}$ calculated
at fixed rigidity, $\vep/Z=75\GeV$. The grammage was calculated for particles with no losses, for a cylindrical
propagation halo with an infinitely thin disc. For the disc grammage, we used $\Xdisc=200\pc\times 1\cm^{-3}\times1.3
m_p$ corresponding to the $200\pc$ wide gaseous disc with $n_\text{H}=0.9\cm^{-3}$ and $n_{\text{He}}=0.1\cm^{-3}$. The
halo geometry, $R=20\kpc$ with a solar location at $r_{s}=8\kpc$, and the diffusion coefficient of
Eq.\eqref{eq:Maurin01D} were the same as in \citep{Maurin01}. For simplicity, the source distribution was taken constant
in the disc.
 The grammage for such a configuration, with a given value of $r_{s}/R$, can be written as
\begin{equation}\label{eq:Xmau}
\Xesc=\Xdisc Lc/(2D)g(L/R)\propto \vep^{-\de},
\end{equation}
where $g(L/R)$ parameterizes the deviation from the one dimensional model and is given by [using \eqref{eq:XdiffHomA}]
\begin{equation}\label{eq:XdiffHom}
g(L/R)=\frac{2R}{L}\sum_{k=1}^\infty J_0\left(\nu_k\frac{r_{s}}{R}\right)
\frac{\tanh\left(\nu_k\frac{L}{R}\right)}{\nu_k^2J_1(\nu_k)},
\end{equation}
where $J_m$ are the bessel functions of the first kind of order $m$ and $\nu_k$ are the zeros of $J_0$.
Eq.~\eqref{eq:Xmau} can be written as
\begin{equation}
K_0/L=\frac{c\Xdisc}{2\Xesc}\frac{g(L/R)}{(\vep/\GeV)^\de},
\end{equation}
reproducing Eq.~\eqref{eq:Maurin01Corr} with $\tilde g\propto g$. As can be seen in figure \ref{fig:Mourin01Fig7} the
reported correlation, Eq.~\eqref{eq:Maurin01Corr}, is explained. The non trivial behavior of $\tilde g(L)$  reflects the
fact that the radial extent of the cylinder cannot be neglected in the calculation for $L\gtrsim 5\kpc$. For smaller
scale heights, the grammage is equal to the expression in Eq. \eqref{eq:XDiffDH}. The function $f(\de)$, normalized to
unity at $\de=0.6$, is given by $f(\de)=(\vep/\GeV)^{\de-0.6}\approx 75^{\de-0.6}$. This is in good agreement with the
required values of $f(\de)$ as can be seen in figure \ref{fig:MaurinFOfDelta}.

As can be seen in figure \ref{fig:Mourin01Fig7}, the allowed parameters have a grammage of
$\Xesc(\vep/Z=75\GeV)=3.5\times(1\pm 0.1) ~\gr\cm^{-2}$ in agreement with the value expected using Eq.~\eqref{eq:X},
$\Xesc(\vep/Z=75\GeV)\approx 3.2 ~\gr\cm^{-2}$.

\subsection{Decay and energy loss suppression $f_{s}(\vep)$ in common models}\label{sec:lossesmodels}
The calculation of $f_{s,e^+}(\vep)$ for the above models is reviewed in \sref{sec:models}. The results, along with
order-of-magnitude estimates are given here. It is useful to parameterize the number density of CR specie $i$ as
\citep{Berezinskii90}
\begin{equation}\label{eq:qtauV}
n_{i}=\tilde q_i\tau_{\text{eff},i}/V_{\text{eff},i},
\end{equation}
where $V_{\text{eff},i}$ is an effective volume probed by the CRs, $\tilde q_i$ is the total production rate in the
effective volume (net production in case of particles experiencing spallation) and $\tau_{\text{eff},i}$ is an effective
production time. In general, energy losses and decay affect both $\tau_{\text{eff},i}(\vep)$, as the losses limit the
available time for particle accumulation, and $V_{\text{eff},i}(\vep)$, which is limited by the distance covered by
particles before they decay or lose a significant part of their energy.

\subsubsection{Leaky Box Model}
In the leaky box model, $\tilde Q_i$ is uniform and thus $\tilde q_i/V_{\text{eff,i}}=\tilde Q_i$ is not affected by
losses. The production time scale is $\taues\propto X_{\esc}$ in the limit of no losses, and approaches $\tau=\taucool
(\tau_d)$ in the opposite limit of significant energy losses (decay). As a result, we have
\begin{equation}\label{eq:floseLBM}
f_{s}(\vep)\sim \tau(\vep)/\taues(\vep).
\end{equation}
For decay, this equation is exact with $\tau=\tau_d$. For energy losses, the relation is [see Eq. \eqref{eq:LBMfs2}]
\begin{equation}
f_{s,e^+}(\vep)=\inv{\gamma-2}\taucool(\vep)/\taues(\vep).
\end{equation}
where $\gamma$ is the observable spectral index. The full solution allowing for moderate losses is given in
\sref{sec:models}. As can be seen, the suppression factor for decaying particles with a decay time $\tau_d$ is similar to
that of positrons with a cooling time $\tau_c$ if $\tau_c\sim\tau_d$. This provides an illustration for the general
argument given in \sref{sec:lossesTheory}.

\subsubsection{Disc halo diffusion}
In the one dimensional disc-halo model, the effective volume should be replaced by an effective scale height
$L_{\text{eff}}$. In the case of no losses, the effective height is simply the halo size, while the effective generation
time is roughly the escape time $\taues$ defined by the relation
\begin{equation}\label{eq:Difftesc}
L\sim \sqrt{D\taues}.
\end{equation}
In the case of significant energy losses (decay), the effective scale height is the distance that CRs propagate before
losing a significant part of their energy (decaying) and is given by
\begin{equation}
L_{\text{eff}}\sim \sqrt{D\tau},
\end{equation}
while the effective generation time is simply $\tau=\tau_c(\tau_d)$. Using Eq. \eqref{eq:qtauV}, we thus have
\begin{equation}\label{eq:floseDiff1}
f_{s}\sim \frac{\tau/\sqrt{D\tau}}{\taues/\sqrt{D\taues}}\sim \sqrt{\tau/\taues}.
\end{equation}
It is useful to parameterize $f_{s}$ by:
\begin{equation}\label{eq:floseDiff2}
f_{s}=C_{\text{Diff}}\frac{\sqrt{D\tau}}{L},
\end{equation}
where $C_{\text{Diff}}$ is a dimensionless coefficient that depends weakly on the injection spectrum and on the
functional form of the diffusion coefficient. As shown in \sref{sec:models}, for decaying particles $C_{\text{Diff}}=1$.
For positrons with a power law injection $Q(\vep)\propto \vep^{-\gamma_i}$ and a diffusion coefficient that depends on
energy as $D\propto \vep^{\de}\oso \Xesc\propto \vep^{-\de}$, $C_{\text{Diff}}$ is given by Eq. \eqref{eq:CDiff}. In the
range $\gamma_i=2.5-3$ and $\de=0.3-0.7$, the values of $C_{\text{Diff}}$ vary between $0.7$ and $0.9$.

Using equations \eqref{eq:floseDiff1},\eqref{eq:posLossICSyn} and \eqref{eq:floseLBM}, we see that the dependence of the
cooling suppression factor $f_{s,e^+}$ on energy is $f_{s,e^+}(\vep)\propto \vep^{(\de-1)/2},\vep^{\de-1}$ for the
disc-halo and Leaky Box models respectively.

\begin{figure}\hspace{-1.3cm}
\epsscale{1.2} \plotone{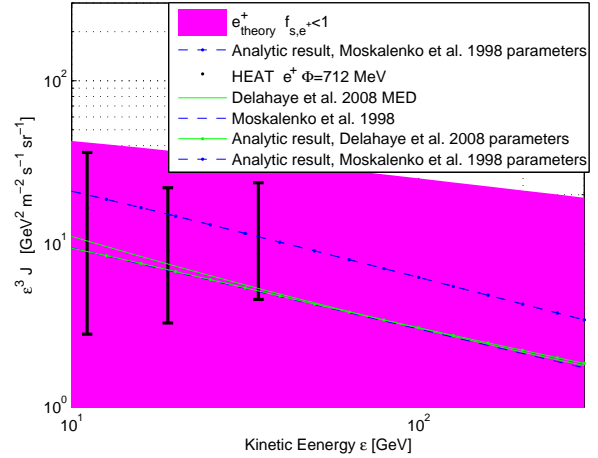}  \caption{The interstellar $e^+$ flux in diffusion models. The filled region is
the expected $e^+$ flux in case there are no losses as in fig \ref{fig:epose}. The solid green and dashed blue lines are
the results of the models in \citep{Delahaye08} [with the MED propagation parameters] and \citep{Moskalenko98}  [as
parameterized by \citep{Baltz99}] respectively.} \label{fig:JPosVsDiff}
\end{figure}

In figure \ref{fig:JPosVsDiff} we compare the results of the $e^+$ flux, with the loss given by equation
\eqref{eq:floseDiff2}, to results of calculations given by \citet{Delahaye08} and \citet{Moskalenko98}. We used equations
\eqref{eq:Jpos0}, \eqref{eq:floseDiff2} with the $e^+$ generation cross sections, the proton spectra, the disc height and
the diffusion coefficient adopted from these references. For the comparison to \citep{Delahaye08} we calculated the
grammage using Eq. \eqref{eq:XDiffDH}, with $\Xdisc=200\pc\times 1\cm^{-3}\times 1.3 m_p$. For comparison with the
results of \citet{Moskalenko98} we used $\Xesc=8.7(\vep/10\GeV Z)^{-0.6}$ normalized to be equal to the value of Eq.
\eqref{eq:X} at $\vep/Z=10\GeV$ and with the power law index taken from \citep{Moskalenko98}.

As can be seen in figure \ref{fig:JPosVsDiff}, our calculation agrees well with that of \citep{Delahaye09} and up to a
factor of $\sim 2$ with that of \citep{Moskalenko98}. We note that the good agreement between the calculations of
\citet{Delahaye09} and \citet{Moskalenko98} is necessarily somewhat accidental, as the cross sections for producing
positrons used by \citet{Moskalenko98} are larger than those used by \citep{Delahaye09} by about $40\%$.

\begin{figure}\hspace{-1.5cm}
\epsscale{1.2} \plotone{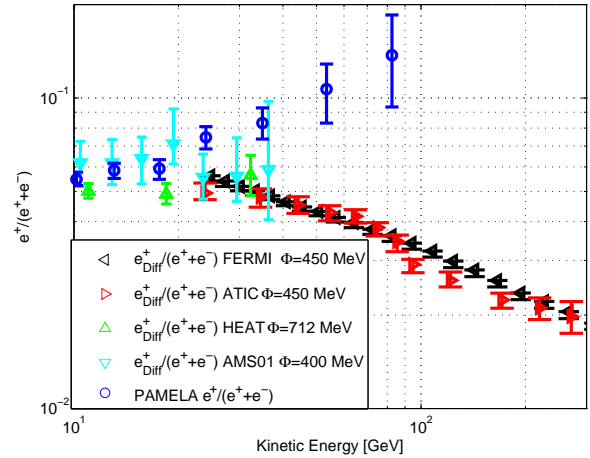}  \caption{The positron fraction as measured by PAMELA compared to the
expectation for diffusion models. The expected $e^+$ flux in the model is divided by the $e^++e^-$ measurements. The loss
suppression is taken to be $f_{s,e^+}=0.33(\vep/10 \GeV)^{-.25}$ corresponding to diffusion with $D\propto \vep^{0.5}$
[see \eqref{eq:floseDiff2}] and normalized to agree with the measurements at $10\GeV$. The observational data is the same
as in fig \ref{fig:PAMELA}.}\label{fig:PamelaVsDiff}
\end{figure}

We next compare the results of diffusion models with the PAMELA measurements using the $e^+ + e^-$ measurements as done
in \sref{sec:positrons}. We assume a suppression factor $f_{s,e^+}=0.33(\vep/10 \GeV)^{-.25}$ with the index
corresponding to $\delta=0.5$ and normalized to agree with the $e^+$ measurements and the estimate based on radioactive
nuclei at $\vep\sim 20\GeV$. The expected $e^+$ flux, divided by $e^++e^-$ measurements is compared to the results of
PAMELA in figure \ref{fig:PamelaVsDiff}.

As can be seen, the expected flux based on this model disagrees with the PAMELA result. We emphasize that this inference
does not depend on the precise amplitude of the $e^+$ production cross sections, or the exact value of $\delta$. This
implies that if PAMELA is correct, these simple diffusion models are ruled out.

The fact that the suppression factor is decreasing with energy is based on the fact that, in these models, the escape
time $\taues\propto \vep^{-\de}$ decreases with energy slower than the cooling time $\tau_c\propto \vep^{-1}$. To see
this, note that the escape time is related to the grammage in these models by
\begin{equation}\label{eq:DiffXvsTvsL}
\Xesc\sim \Xdisc c\taues/L,
\end{equation}
where we used Eqs. \eqref{eq:XDiffDH} and \eqref{eq:Difftesc}. This has a simple interpretation - the distance a particle
travels before escaping divided by the confining scale is similar to the number of disk crossings. Using
\eqref{eq:DiffXvsTvsL}, the fact that the grammage $\Xesc$ changes as $\vep^{-\de}$ and that $L$ is assumed to be
constant,  we see that $\taues$ changes as $\taues\propto \vep^{-\de}$.

A crucial assumption here, that may be wrong, is that $L$ does not depend on energy \citep[e.g.][]{Dgoel93}. In fact, a
scale height decreasing with energy as $L(\vep)\propto \vep^{-\delta_L}$ would imply an escape time changing as
$\taues\propto\vep^{-\de-\delta_L}$ that may be declining faster than $\tau_c$ if $\delta_L>1-\de$. Note that to a good
approximation, the estimate for the anti-proton and positron fluxes can be obtained in such a case by replacing $L$ in
equations \eqref{eq:XDiffDH},\eqref{eq:floseDiff2} and \eqref{eq:floseDiffA} with $L(\vep)$. To see this, note that
anti-protons and CR nuclei do not change their energy during propagation, implying that these equations are directly
applicable, while positrons and radioactive nuclei are not affected by the boundary conditions as long as they are
strongly suppressed by losses. The expected positron fluxes are affected by the energy dependence of L indirectly,
through the energy dependence of $D(\vep)$ inferred from the grammage measurements.

Alternatively, $\tau_c$ may deviate from the $\vep^{-1}$ dependence if the energy losses strongly depend on the position
in the halo.

\section{Claims that a primary positron source is necessary}
\label{sec:claim_prim}

The claims made in the literature, that the PAMELA data requires a primary $e^+$ source
\citep[e.g.][]{Adriani09a,Morselli08}, are based on one of two different lines of reasonings:
\begin{itemize}
\item The electrons are assumed to have the same production spectrum as the protons, and to suffer the same energy losses as the positrons $f_{s,e^-}=f_{s,e^+}$. Under these assumptions, the $e^-$ spectrum is given (up to an energy independent factor) by
\begin{equation}
n_{e^-}(\vep)\propto f_{s,e^+}(\vep)n_p(\vep),
\end{equation}
while the $e^+$ spectrum is given (up to a factor that depends weakly on energy) by
\begin{equation}
n_{e^+}(\vep)\propto f_{s,e^+}\Xesc(\vep/Z)n_p(\vep),
\end{equation}
resulting in a $e^+$ to $e^-$ ratio,
\begin{equation}
n_{e^+}(\vep)/n_{e^-}(\vep)\propto \Xesc(\vep/Z)\propto \vep^{-0.5},
\end{equation}
that is independent of the unknown energy loss term. The observed slightly rising positron fraction is in contradiction
with this expectation and thus a primary source is claimed to be necessary. This argument does not rely on direct $e^-$
measurements.

\item The $e^+$ flux, including the energy loss suppression, is calculated within a specific propagation model. The prediction of
the model is divided by the measured $e^-+e^+$ flux and claimed to disagree with the positron fraction [as discussed in
\sref{sec:lossesmodels}, see fig \ref{fig:PamelaVsDiff}].
\end{itemize}

The first line of argument relies on two unestablished assumptions, namely that the electrons have the same injection
spectrum as the protons and that electrons and positrons suffer the same cooling suppression. The acceleration of
electrons and their escape from the sources are far from being understood. Even if CRs are accelerated in SNRs, as
commonly assumed, and the electrons are accelerated with a spectrum similar to protons at a given shock velocity, the
integrated spectrum of electrons leaving the SNR may be non trivial, due to energy losses and possible different
acceleration efficiencies at different shock velocities. In addition, the electrons may suffer different losses than
positrons on their way to earth as their sources are distributed differently then the $e^+$ sources
\citep[e.g.][]{Shaviv09,Piran09}.

The second line of argument relies on the details of the specific propagation model that is being used (see discussion in
\sref{sec:OtherModels}). The declining suppression factor, which is in contradiction with the PAMELA measurements,
results from the fact that the escape time drops with energy faster than the cooling time in the considered models. Such
a relation between the escape and cooling times is not based on observations and can be modified in alternative models
(see \sref{sec:lossesmodels}).

\section{Previously suggested models with a secondary origin for the positrons}\label{sec:SecModels}

In \sref{sec:simple} and \sref{sec:losses} it was shown that the $e^+$ and $\bar p$ measurements are consistent with
model independent constraints of a secondary origin for these particles. However, when the commonly used assumptions
regarding the $e^-$ and $e^+$ generation and propagation described in \S~\ref{sec:claim_prim} are adopted, a conflict
with the PAMELA results arises. In this section we discuss previously suggested specific models, which do not adopt the
assumptions described in \S~\ref{sec:claim_prim}, and which were constructed to reproduce the PAMELA measurements with
positrons being of secondary origin.

In \citep{Blasi09,Blasi09b,Mertsch09} it was suggested that the secondary positrons are mainly generated inside SNRs
resulting with a positron fraction that is flatter than in the usual picture of secondary CR production. To estimate the
contribution of SNRs, the amount of grammage traversed by the primary particles during the lifetime of the SNRs should be
compared to the grammage traversed in the Galaxy. The grammage traversed in the SNRs is approximately given by
\begin{align}
X_{\text{SNR}}&\sim 4t_{\text{SNR}}c \rho_{\text{ISM}}\cr &\sim
0.08\left(\frac{t_{\text{SNR}}}{10^4\yr}\right)\frac{n_{\text{ISM}}}{1\cm^{-3}}\gr\cm^{-2},
\end{align}
where $t_{\text{SNR}}\sim 10^4\yr$ is the age of the SNR and $4\rho_{\text{ISM}}$ is approximately the post-shock
density. This should be compared to the grammage traversed by the primary protons at $\vep_p\sim 10\vep$ through the
Galaxy given by
\begin{align}
X_{\text{Gal}}\sim 3 (\vep_p/100\GeV)^{-0.5}\gr\cm^{-2}.
\end{align}
As can be seen, the contribution of anti-protons and positrons generated in SNRs is negligible for $100\GeV<
\vep_p<300\GeV$, where the Galactic grammage is measured. This is illustrated in \citep{Blasi09b} for anti-protons in the
corresponding energy range $10\GeV<\vep<30\GeV$. In disk+halo models, the relative contribution of the SNRs, compared to
the galactic contributions for positrons and anti-protons is the same, regardless of the different cross sections or the
losses. In fact, the relative contribution is equal to the ratio of the SNR secondary production rate to that in the ISM.
It is therefor not clear how the positrons at $\vep\sim 10\GeV$ turn out to be dominated by the contribution from SNRs in
\citep{Blasi09}.

The effect of acceleration of these secondaries inside the SNRs, assuming diffusive shock acceleration theory, is limited
to the secondaries produced during the acceleration time. For these, the acceleration amplifies the density at a given
energy by a factor of a few (acceleration of a $\vep^{-2}$ spectrum results in a $\vep^{-2}$ spectrum, somewhat
upshifted). This is an order unity effect, relevant only at the highest energies, where the acceleration time is similar
to the SNR age, and where the primary flux begins to cut off.

We note that at $\vep_p\gtrsim 1\TeV$, corresponding to secondary anti-protons and positrons at $\vep\gtrsim 100\GeV$,
the contribution depends on unmeasured grammage and there is more room for an SNR contribution. In addition, if SNRs
explode in dense, $n_{\text{ISM}}\gg 1\cm^{-3}$ regions, the contribution may be important. In any case, any such effect
should be observed for positrons and anti-protons at roughly the same energy, and for nuclei spallation products at
higher energies.

In \citep{Shaviv09,Piran09} it was suggested that primary electrons are generated in a nonuniform distribution of sources
that follow the spiral arms, while positrons are generated throughout the gaseous disc. This is an interesting example
for the effect of a non trivial primary source distribution resulting in a difference in the effects of energy losses of
electrons and positrons. However, the $e^-$ flux is constructed in this model to agree with the $e^++e^-$ measurements,
while the $e^+$ are produced roughly uniformly in the disc, similarly to common diffusion models. It is therefore
unclear, how the positron fraction obtained in this model is different than that of ordinary diffusion models, which are
fitted to the same $e^++e^-$ measurements and in which the positrons are generated and propagated in roughly the same way
(see \sref{sec:lossesmodels}).

Note that this model was not confronted with compositional CR data. In particular, it is expected that the heavy primary
element contribution from the spiral arms will be exponentially suppressed due to spallation losses in this model, in
contrast to the observed abundances.

\citet{Cowisk09a,Cowisk09b} outlined a model in which at low energies particles spend most of their energy dependent time
in the vicinity of the sources before escaping into the Galaxy, where the residence time is energy independent. In this
model, the primary observed spectra are the same as the source injection spectra and the accumulated grammage flattens
off with energy above a threshold energy  $\vep_t$.  By choosing appropriate parameters, the model can account for the
observed positron fraction while being consistent with nuclei spallation products at $\vep\lesssim 10\GeV/\text{nuc}$.
This model seems to disagree with secondary to primary ratio measurements at higher energies
[\citep{Engelman90,Binns88,Ahn08}, see however \citep{Zatespin09}]. A break in the $\bar p$ spectrum at $\sim 10\GeV$ is
expected in such a model. Thus it is crucial to compare the prediction of these models with $\bar p$ observations.

A feature common to the above model and to the model suggested by \citep{Blasi09}, is that the spectrum of the primary
CRs in the regions where the secondaries are produced is different than the spectrum at the solar system. These are
interesting examples where deviations from the expressions for the $e^+$ and $\bar p$ fluxes \eqref{eq:PbarOverP} and
\eqref{eq:Jpos0} are expected, due to the fact that $\Xesc$ is determined by observations of spallation products which
are generated at rigidities similar to their primaries, while the positrons and anti-protons are generated at lower
energies. For these models, the appropriate grammage to consider is the grammage evaluated at the primary energy. Note,
that the ratio of $e^+$ to $\bar p$ fluxes Eq. \eqref{eq:epbar}, is insensitive to such spectral variations of the
primary CRs, as illustrated in figure \ref{fig:epbar}.

\section{Summary and Discussion}\label{sec:Discussion}
In this paper we presented a critical analysis of the observational constraints on, and the theoretical modeling of,
aspects of CR generation and propagation in the Galaxy, which are relevant for the interpretation of recent $e^+$ and
$\bar p$ measurements.

In \sref{sec:Grammage} we reviewed the main observationally established properties of CR propagation. We focused on the
analysis of stable CR nuclei measurements and argued that they can be accurately described by simple equations,
\eqref{eq:SecondaryRatioEquation}-\eqref{eq:Grammage2} that are valid for a broad range of models, namely models which
satisfy the conditions \eqref{eq:Conditions}. The basic principle behind these equations is that the CR density should be
proportional to the CR generation rate as long as the CR composition is uniform. The main challenge in the analysis of
secondary CRs, including anti-protons and positrons, comes from the fact that CRs suffer during their propagation
significant losses, which are different for different species. Nuclei suffer spallation and decay, while positrons suffer
energy losses due to Inverse Compton and Synchrotron emission. The spallation losses of stable nuclei can be absorbed
into a net production rate, see Eq.~\eqref{eq:GenerationRate}.

The CR grammage $\Xesc$, see Eq.~\eqref{eq:X}, extracted from the stable CR nuclei measurements, allowed us to write down
in \sref{sec:simple} analytic, model independent constraints for the expected secondary $\bar p$ flux \citep[a la][
equivalent to Leaky Box and thin-disc+halo models]{Gaisser92,Simon98}, and for the secondary $e^+$ flux, including an
accurate estimate of the $\bar p$ flux, Eq. \eqref{eq:PbarOverP}, and an upper limit for the $e^+$ flux,
Eq.~\eqref{eq:Jpos}, obtained by neglecting the $e^+$ radiative losses. We have shown (\sref{sec:OtherModels}) that the
simple analytic expressions reproduce the results of detailed calculations, based on popular models for CR propagation,
including the leaky box model and disc+halo diffusion models \citep[e.g.][]{Maurin01,Delahaye08,Moskalenko98}.
Anti-proton and $e^+$ measurements available up to $\vep\sim 100$ GeV, including the PAMELA positron fraction
measurement, were shown to be consistent with our analytic estimates in figures
\ref{fig:PbarOverP}-\ref{fig:PAMELAKamae}.

The claims that the measured PAMELA positron fraction requires primary $e^+$ sources were shown in
\S~\ref{sec:claim_prim} to be based on assumptions, that are not supported by observations.  If PAMELA results are
correct, the fact that they are in disagreement with the results of models based on these assumptions simply imply that
the assumptions are not valid. In particular, the models which are ruled out include (a) models in which the primary
electrons are injected with a source spectrum equal to that of protons and in which the electrons and positrons suffer
identical losses, and (b) thin-disc+halo diffusion models, with an isotropic diffusion equation with energy independent
halo size.

Positron flux measurements may be used to infer the suppression factor $f_{s,e^+}(\vep)$ [see Eq. \eqref{eq:fDef}] due to
radiative energy losses. Existing $e^-$ and $e^+$ data were shown in \S~\ref{sec:simple} to imply that
$f_{s,e^+}(\vep)\sim 1/\text{few}$ in the PAMELA energy range, and that $f_{s,e^+}(\vep)$ is indicated to be slightly
growing with energy, see figure \ref{fig:PAMELA} and \ref{fig:PAMELAKamae} and the discussion following them. In
\sref{sec:losses} we showed that measurements of radioactive unstable secondaries allow an indirect estimate of
$f_{s,e^+}\sim1/3$ at a single energy, $\sim20\GeV$ (\sref{sec:lossesAp}), consistent with the $e^+$ measurements. The
$e^+$ measurements at higher energies, $\vep\gtrsim 20$ GeV, provide new information on the temporal behavior of CR
propagation that is not currently accessible by other types of experiments. Measurements of the $e^+$ to $\bar p$ flux
ratio are particularly useful for this purpose, as well as for challenging secondary production models, as this ratio is
insensitive to model details (see figure \ref{fig:epbar}).

In section \sref{sec:SecModels} we discussed specific models, in which the positrons are of secondary origin, that were
suggested to explain recent $e^+$ abundance measurements \citep{Blasi09,Shaviv09,Cowisk09a}. We showed that these models
appear to be in conflict with constraints derived from CR measurements. It is important to emphasize, that there is a
variety of models that can be constructed, in which both $e^+$ and anti-protons are of secondary origin, and which are
consistent with the constraints arising from CR observations. The main point that should be addressed in such models is
the energy dependence of the $e^+$ suppression factor $f_{s,e^+}$ at energies exceeding $\vep\gtrsim 20\GeV$, which is
indicated by the PAMELA measurements to be slightly rising with energy. One example, that could result in such a
behavior, is to have energy dependent boundary conditions to the transport equation \citep[e.g.][see
\sref{sec:lossesmodels}]{Dgoel93}.

% -------------------------- End of Discussion --------------------------

\acknowledgments
%This research was partially supported by ISF, AEC and Minerva grants.
We thank W. R. Webber, T. Volansky and P. Salati for useful communications. This research was supported by ISF, AEC \&
Minerva grants.

%-----------------------------------------------------------------------------
% --------------------------      BIBLIOGRAPGHY ---------------------------
%-----------------------------------------------------------------------------

%\clearpage
\appendix
\section{CR propagation models}\label{sec:models}
Below we provide the calculations quoted in \sref{sec:OtherModels}. Calculations along similar lines can
be found in the literature \citep[e.g.][]{Ginzburg76,Berezinskii90}. Nevertheless, we find it convenient to attach in full the treatment of specific problems encountered in the paper.

\subsection{Cooling suppression of positrons in the Leaky Box Model (LBM)}
For positrons, the LBM equation reads
\begin{equation}\label{eq:posLBM}
Q_{e^+}=\frac{n_{e^+}}{\tau_{\esc}}+\pr_{\vep}(\dot\vep n_{e^+})
\end{equation}
where $-\dot\vep=\vep/\taucool \propto\vep^2$ is the energy loss rate. Solving Eq.~\eqref{eq:posLBM} we obtain the suppression factor due to cooling,
\begin{equation}\label{eq:posLBM2}
f_{s,e^+}(\vep)=\frac{\taucool (\vep)}{\tauesc (\vep)} \frac{\int_\vep^\infty d\vep'Q_{e^+}(\vep')\exp\left[-\int_\vep^{\vep'}\frac{d\vep''\taucool (\vep'')}{\vep''\tauesc (\vep'')}\right]}{\vep Q_{e^+}(\vep)}.
\end{equation}
In the limit $\taucool\ll\tau_{\esc}$, the exponential term can be omitted and we get
\begin{equation}\label{eq:LBMfs}
f_{s,e^+}(\vep)\approx\frac{\taucool (\vep)}{\tauesc (\vep)}\frac{\int_1^{\infty}d\xi
Q_{e^+}(\xi\vep)}{Q_{e^+}(\vep)}=\frac{1}{\gamma_i-1}\frac{\taucool }{\tauesc },
\end{equation}
where the last equality holds for a power law injection spectrum, $Q_{e^+}(\vep)\propto \vep^{-\gamma_i}$. In the limit
$\taucool\ll\tau_{\esc}$, the injection index $\gamma_i$ is related to the observed index $\gamma$ by $\gamma_i=\gamma-1$
and so Eq. \eqref{eq:LBMfs} can be written as:
\begin{equation}\label{eq:LBMfs2}
f_{s,e^+}\approx\frac{1}{\gamma-2}\frac{\taucool }{\tauesc }.
\end{equation}

\subsection{Diffusion model with Halo and thin disc}
\paragraph{Stable secondaries with no energy losses, cylindrical model}
For the case considered in \sref{sec:OtherModels}, it is useful to perform a bessel decomposition in the
radial coordinate while solving for the $z$ coordinate directly. The solution for the CR density is given by
\begin{equation}\label{eq:ng}
n(z,r;\vep)=\frac{R}{D(\vep)}\sum_{k=1}^\infty J_0\left(\nu_k\frac{r}{R}\right)\frac{q_k(\vep)}{2\nu_k}\left[\tanh\left(\nu_k\frac{L}{R}\right)\cosh\left(\nu_k\frac{z}{R}\right)
-\sinh\left(\nu_k\frac{z}{R}\right)\right],
\end{equation}
where $q_k$ are the coefficients in the bessel decomposition of the source surface density function $q(r;\vep)$, defined
on the gaseous disc. We assume a source function which is uniform across the disk $q(r;\vep)$=$q(\vep)$, which has the
following decomposition
\begin{equation}\label{eq:qg}
q_k(\vep)=\frac{2 q(\vep)}{\nu_kJ_1(\nu_k)}.
\end{equation}
The grammage is defined by the equation $n=\Xesc Q/(\rho_{\text{ISM}}c)$  [Eq. \eqref{eq:GrammageNS}]. Here the local
generation rate per unit mass is given by $Q/\rho_{\text{ISM}}=q(r_{s};\vep)/\Xdisc$. Using \eqref{eq:ng} and
\eqref{eq:qg}, the local density can be expressed as
\begin{equation}
n(z=0,r=r_{s};\vep)=\frac{R\tilde q(\vep)}{D(\vep)}\sum_{k=1}^\infty
J_0\left(\nu_k\frac{r_{s}}{R}\right)\frac{\tanh\left(\nu_k\frac{L}{R}\right)}{\nu_k^2J_1(\nu_k)}.
\end{equation}
The grammage for this model is therefore
\begin{equation}\label{eq:XdiffHomA}
\Xesc=\Xdisc\frac{Lc}{2D}\times\frac{2R}{L}\sum_{k=1}^\infty J_0\left(\nu_k\frac{r_{s}}{R}\right)
\frac{\tanh\left(\nu_k\frac{L}{R}\right)}{\nu_k^2J_1(\nu_k)}.
\end{equation}
In the limit $L\ll R$ the summed factors $\tanh(\nu_kL/R)\to\nu_kL/R$, and the result Eq. \eqref{eq:XDiffDH} of the one
dimensional model is reproduced.

\paragraph{Decaying particles, one dimensional limit}
For decaying particles with observer frame decay time $\tau(\vep)$, the solution to the diffusion equation is
\begin{equation}
n(z;\vep)=\frac{q}{2}\sqrt{\frac{\tau}{D}}\left[\tanh\left(\frac{L}{\sqrt{D\tau}}\right)\cosh\left(\frac{z}{\sqrt{D\tau}}\right)
-\sinh\left(\frac{z}{\sqrt{D\tau}}\right)\right],
\end{equation}
where $q,D,\tau$ depend on $\vep$. The suppression factor defined at $z=0$ is
\begin{equation}
f_{s,i}=\frac{\sqrt{D\tau}}{L}\tanh\left(\frac{L}{\sqrt{D\tau}}\right)\xra[\sqrt{D\tau}\ll L]{}\frac{\sqrt{D\tau}}{L}.
\end{equation}

\paragraph{Continuous energy losses, one dimensional limit}

We next consider particles that suffer energy dependent cooling $\dot \vep(\vep)$.

It is useful to express the steady state density in terms of the time dependent number density of particles $n(t,\vep)dt$
that would result if the source generated particles at all energies and throughout the disk at an interval $dt$ at time
$t=0$ by
\begin{equation}\label{eq:SteadyStateTime}
n(\vep)=\int_0^{\infty}n(t,\vep)dt.
\end{equation}

It is straight forward to show that
\begin{equation}
n(t,\vep)=q(\vep_S)\frac{d\vep_S}{d\vep}n_0[\tilde t(t,\vep_S)]
\end{equation}
where $\vep_S$ is the energy a particle had at time $t=0$ in order to be at energy $\vep$ at a fictitious time $\tilde t$ defined by
\begin{equation}
\tilde
t(t,\vep_S)=\int_0^tD[\vep(\vep_S,t')]dt'=\tau_c(\vep)D(\vep)\int_1^{\vep_S/\vep}\frac{D(\xi\vep)}{D(\vep)}\frac{\dot
\vep(\vep)}{\dot \vep(\xi\vep)}d\xi,
\end{equation}
and $n_0(\tilde t)$ is the value at $z=0$ of the solution to the spatial diffusion equation $\pr_{\tilde t}n_0=\pr_z^2n_0$, with initial distribution $n_0(\tilde t=0,z)=\de(z)$ and a constant diffusion coefficient equal to unity.
Using Eq. \eqref{eq:SteadyStateTime}, the steady state distribution is given by
\begin{equation}
n(\vep)=q(\vep)\tau_c(\vep)\int_1^{\infty}\frac{q(\xi\vep)}{q(\vep)}n_0(\tilde t)d\xi.
\end{equation}
For source and diffusion coefficient with a power law dependence on energy, $q(\vep)\propto \vep^{-\gamma_i}$ and
$D(\vep)\propto \vep^{\de}$, and taking $\dot \vep\propto \vep^{-2}$ we find
\begin{equation}
\tilde t(\vep_S)=\tau_c(\vep)D(\vep)\int_1^{\vep_S/\vep}
\xi^{\de-2}d\xi=\inv{1-\de}[1-(\vep_S/\vep)^{\de-1}]\tau_c(\vep)D(\vep),
\end{equation}
and
\begin{equation}\label{eq:nclossesf}
n=q\tau_c\int_1^{\infty}\xi^{-\gamma_i}n_0\left(\inv{1-\de}(1-\xi^{\de-1})\tau_cD\right)d\xi,
\end{equation}
where all physical quantities ($n,q,D,\tau_c$) are to be evaluated at $\vep$. In the limit of strong suppression we
find
\begin{equation}
n=\frac{q\tau_c}{\sqrt{4\pi D \tau_c}}\sqrt{1-\de}\int_1^{\infty}\xi^{-\gamma_i}(1-\xi^{\de-1})^{-1/2}d\xi,
\end{equation}
in agreement with the result in \citep{Delahaye08}. The implied suppression factor is
\begin{equation}\label{eq:floseDiffA}
f_{s,e^+}=C_{\text{Diff}}\frac{\sqrt{D\tau_c}}{L},
\end{equation}
where $C_{\text{Diff}}$ is a dimensionless coefficient given by
\begin{equation}\label{eq:CDiff}
C_{\text{Diff}}=\sqrt{\frac{1-\de}{\pi}}\int_1^{\infty}\xi^{-\gamma_i}(1-\xi^{\de-1})^{-1/2}d\xi.
\end{equation}
In the range $\gamma_i=2.5-3$ and $\de=0.3-0.7$ the values of this function are between $0.7$ and $0.9$.

\bibliographystyle{apj}

% ------------------------------ End of bibliography --------------------

\end{document}

%% file: Definitions.tex
\newcommand{\DDir}{\relax{D\kern-.7em{/}}}

\newcommand{\inv}[1]{\frac{1}{#1}}

\newcommand{\ra}{\rightarrow}

\newcommand{\xra}{\xrightarrow}

\newcommand{\oso}{\Leftrightarrow}

\newcommand{\be}{\begin{equation}}
\newcommand{\ee}{\end{equation}}
\newcommand{\bea}{\begin{equation*}}
\newcommand{\eea}{\end{equation*}}

\newcommand{\pr}{\partial}

\newcommand{\nin}{\relax{\in\kern-.8em{/}}}

\newcommand{\Te}{\Theta}

\newcommand{\bt}{\beta}

\newcommand{\de}{\delta}

\newcommand{\sig}{\sigma}

\newcommand{\vep}{\varepsilon}

\newcommand{\cm}{\mbox{ cm}}
\newcommand{\m}{\mbox{ m}}
\newcommand{\sr}{\mbox{ sr}}
\newcommand{\se}{\mbox{ s}}
\newcommand{\yr}{\mbox{ yr}}

\newcommand{\mb}{\mbox{ mb}}

\newcommand{\pc}{\mbox{ pc}}
\newcommand{\kpc}{\mbox{ kpc}}

\newcommand{\eV}{\mbox{eV}}

\newcommand{\GeV}{\mbox{~GeV}}
\newcommand{\TeV}{\mbox{ TeV}}

\newcommand{\gr}{\mbox{g}}

\newcommand{\sref}{\S~\ref}